\documentclass[12pt,aps,english]{revtex4}

\usepackage{graphicx}
\usepackage{bm}
\usepackage{dcolumn}
\usepackage[usenames, dvipsnames]{color}
\usepackage{epsfig}
\usepackage{amsmath}

\usepackage{color}

\begin{document}

\title{Entropy production and work fluctuation relations for a single particle in active bath}
\author{Subhasish Chaki and Rajarshi Chakrabarti*}
\affiliation{Department of Chemistry, Indian Institute of Technology Bombay, Powai, Mumbai 400076, E-mail: rajarshi@chem.iitb.ac.in}
\date{\today}

%\documentclass{article}
%%\begin{document}
%\hspace{3 cm} Subhasish Chaki
%\vspace{1 cm} 
\maketitle

\section*{Abstract}
A colloidal particle immersed in a bath of bacteria is a typical example of a passive particle in an active bath. To model this, we take an overdamped harmonically trapped particle subjected to a thermal and a non-equilibrium noise arising from the active bath. The harmonic well can be attributed to a laser trap or to the small amplitude motion of the sedimented colloid at the bottom of the capillary. In the long time, the system reaches a non-equilibrium steady state that can be described by an effective temperature. By adopting this notion of effective temperature, we investigate whether fluctuation relations for entropy hold. In addition, when subjected to a deterministic time dependent drag, we find that transient fluctuation theorem for work cannot be applied in conventional form. However, a steady state fluctuation relation for work emerges out with a renormalized temperature.   

\section{Introduction}
Mesoscopic objects such as colloidal particles, proteins etc. in a fluid medium undergo random motion due to bombardment from the surrounding fluid particles. These are examples of dynamics within equilibrium description and can be described in the framework of Brownian Motion \cite{zwanzig2001nonequilibrium}. However the situation is quite different if the particle is suspended in a bath of active particles such as bacteria \cite{argun2016non, krishnamurthy2016micrometre}. Recently Maggi $et\,\,al.$ experimentally and numerically investigated the dynamics of colloidal beads in a bath of swimming $E.coli$ bacteria \cite{maggi2014generalized}. They found that, collisions from the swimming bacteria result enhanced diffusion of the colloid. In case of a polymer in a bacteria bath, it has been shown theoretically that the MSD of a tagged monomer grows faster and the polymer undergoes swelling \cite{harder2014activity, samanta2016chain, vandebroek2015dynamics,kaiser2014unusual,eisenstecken2016conformational,eisenstecken2017internal, osmanovic2017dynamics}. These are examples of passive particles in active bath. Other examples of active processes include polar pattern formation in driven filament systems \cite{suzuki2015polar}, the motion of the cytoskeleton inside cells controlled by ATP driven motor proteins \cite{brangwynne2008nonequilibrium}, biological membranes constantly maintained out of equilibrium by active proteins inside the membrane  \cite{faris2009membrane} etc. Getting inspired by these active processes, attempts have been made to develop theories on model systems such as a single colloid or a single polymer in an active bath \cite{harder2014activity,samanta2016chain, vandebroek2015dynamics,kaiser2014unusual,eisenstecken2016conformational,eisenstecken2017internal, osmanovic2017dynamics,shin2015facilitation,shin2017elasticity}. In a very recent work, Clausius inequality for active particles has been proposed \cite{marconi2017,marconi2017heat}. It is obvious that the presence of active particles drives the system also away from equilibrium and cannot be described using equilibrium theories of passive Brownian motion. 

In general, fluctuation relations    \cite{sevick2008fluctuation,ritort2008nonequilibrium,martinez2017colloidal,
varghese2013force,colangeli2011steady,saha2011work,saha2007asymmetry} hold for the mesoscopic systems driven away from thermal equilibrium. In such cases, work, heat and entropy are fluctuating quantities and that may lead to an apparent violation of second law of thermodynamics. However the second law of thermodynamics is recovered by taking averages over many trajectories. Considering a system which is intially at thermal equilibrium and then driven away by an external force for a finite time interval,  transient fluctuation theorem (TFT) states $\frac{P_F(S)}{P_R(-S)}=e^{\frac{S}{k_B}}$  ,where $P_F(S)$ and $P_R(S)$ are the probabilities of entropy production in forward and time reversed processes respectively   \cite{evans1994equilibrium,kurchan1998fluctuation,crooks1999entropy}. If the system is already in a non-equilibrium steady state throughout  under constant driving, then steady-state fluctuation theorem (SSFT) holds {\cite{gallavotti1995dynamical,speck2005integral}. In all these situations, entropy production is associated to the heat bath and there is an unique ambient temperature $T$, the temperature of the bath \cite{kurchan1998fluctuation,van2003extension}. If one includes the entropy change of the system, then total entropy production along a single trajectory follow the integral fluctuation theorem (IFT), $\left<e^{-\frac{\Delta S_{tot}}{k_B}}\right>=1$ \cite{seifert2005entropy}. Experimentally these relations have been extensively verified on colloidal particle driven by a constant force along a periodic potential \cite{speck2007distribution}, the circuit of an electric dipole in electric potential bias \cite{garnier2005nonequilibrium}, and a single molecule of RNA under mechanical stretch \cite{hummer2001free, liphardt2002equilibrium}. On the other hand, it has been shown both analytically and numerically that fluctuation relations can not be applied in some cases \cite{beck2004superstatistical,touchette2007fluctuation,sellitto2009fluctuation,
chechkin2009fluctuation,harris2009current,budini2012generalized}. However for both glassy and Gaussian stochastic dynamics, the functional form of fluctuation relation has been recovered by replacing the ambient thermal temperature $T$ with the nonequilibrium effective temperature \cite{zamponi2005generalized,zamponi2005fluctuation,chechkin2012normal}. 

The immediate question arises whether the fluctuation relations can also be applied to these active systems. Along this direction, very recently, the fluctuation relations and stochastic thermodynamics of active systems such as single enzymes and molecular motors have gained much attention \cite{seifert2011stochastic,lacoste2009fluctuation,speck2016stochastic}. In addition, entropy production has also been investigated for active brownian particles \cite{chaudhuri2014active,pietzonka2017entropy,mandal2017entropy}. In a very recent experiment Argun $et\,\,al.$ \cite{argun2016non} have shown that Crooks fluctuation theorem \cite{crooks1999entropy}, the Jarzynski equality \cite{jarzynski1997nonequilibrium, jarzynski1997equilibrium}, and the integral fluctuation theorem \cite{seifert2005entropy}  cannot be applied to active baths. They also showed that if the trap relaxation time is comparable to or shorter than the characteristic time scale associated with the active noise, then non-Gaussian statistics emerges. Independently, Krishnamurthy $et\,\,al.$ have shown that the displacement statistics of a colloidal particle in a time-varying optical potential across bacterial baths becomes increasingly non-Gaussian with the activity of the bacterial bath \cite{krishnamurthy2016micrometre}. However at very low bacteria concentration, the dynamics of the colloidal particle is expected to be Gaussian \cite{maggi2014generalized,wu2000particle}. These active fluctuations arising due to motion of bacteria or molecular motors, have been theoretically modeled using Gaussian random variable with zero mean and an exponentially decaying temporal correlation \cite{vandebroek2017effect, vandebroek2015dynamics, berthier2013non}. In addition, it has been experimentally shown that the displacement of a tracer bead immersed in a actomyosin network has a Gaussian distribution superimposed with fat exponential tails \cite{stuhrmann2012nonequilibrium, toyota2011non}. For low myosin concentrations the distribution is purely Gaussian \cite{sonn2017scale}.     
 
In this paper, we deal with an exactly solvable model of the dynamics of a passive tracer in a harmonic well coupled to a thermal and an active bath. In case of  weak trapping and high viscous medium such as inside a biological cell, the characteristic time scale of the harmonic trap is longer than the correlation time of the active noise, then the process can be considered as Gaussian and our model would fit in. The presence of Gaussian active noise then can be attributed to an effective temperature as done in the previous studies \cite{samanta2016chain,ghosh2014dynamics,vandebroek2015dynamics}. Here we check the validity of transient and steady state work fluctuation theorem for a dragged harmonic oscillator in the presence of a Gaussian active noise. We also analyze entropy production and find that the fluctuation relation for entropy production cannot be applied. On the other hand, by invoking the notion of an effective temperature, we show that the transient fluctuation theorem (TFT) for work cannot be applied in this case. But a steady state fluctuation theorem for work, emerges out with a renormalized inverse temperature $(\alpha)$ different from $\beta\left(=\frac{1}{k_BT}\right)$. 

\section{Entropy production in active bath without external time dependent protocol}
We consider a Brownian particle in a one dimensional harmonic well in contact with a heat bath at temperature $T$. The harmonic well mimics the laser trap as used by Argun $et\,\,al. $\cite{argun2016non} or in a different experimental set up, accounts for the small amplitude motion of the sedimented colloid at the bottom of the capillary \cite{maggi2014generalized}. The effects of active bath enters in the particle's motion through an extra noise, $\eta_A(t)$. Active Bio-systems are associated with low Reynolds numbers for which $\frac{m}{\gamma} \rightarrow 0$ \cite{vandebroek2017effect}. 

  So the dynamics is best described by an overdamped Langevin equation 
\begin{equation}
\gamma \frac{dx}{dt} = -kx + \xi(t) + \eta_A (t)
\label{eq:langevinact}
\end{equation}
Where $\gamma$ is the friction coefficient and $k$ is the spring constant for the harmonic trap. $\xi(t)$ is the Gaussian thermal noise with the statistical properties \cite{zwanzig2001nonequilibrium}
\begin{equation}
\left<\xi(t)\right>=0, \left<\xi(t) \xi(t^\prime)\right> = 2 \gamma k_B T \delta(t-t^\prime)
\end{equation}
The statistical properties of $\eta_A(t)$ which also has a Gaussian distribution are
\begin{equation}
\left<\eta_A(t)\right>=0,
\left\langle \eta_A(t) \eta_A(t^{\prime}) \right\rangle=Ce^{-\frac{|t-t^{\prime}|}{\tau_A}}
\end{equation}
$C$ refers to the strength of the of the active noise and $\tau_a$ is the persistence time of the bacterial forces acting on the particle associated with a persistence length $L_a$ and $C=\frac{L_{a}^2}{\tau_{a}^2}$ \cite{argun2016non}. Here the active noise $\eta_A(t)$ does not follow any fluctuation-dissipation relation \cite{samanta2016chain,vandebroek2015dynamics}. This exponential correlation is a reminiscent of active dynamics such as Run-and-Tumble
particles, active Brownian particles, and active Ornstein–Uhlenbeck motion \cite{zakine2017stochastic}.  In case of a weak trapping and highly viscous medium, the trap relaxation time $\left(\frac{\gamma}{k}\right)$ is longer than the bacterial correlation time $\tau_A$ causing a complete separation of the time scales. This allows us to model the active noise, $\eta_A(t)$ as a Gaussian random variable \cite{argun2016non,kanazawa2012stochastic,krishnamurthy2016micrometre}. In addition, a series of theoretical studies have been performed by treating the active noise as a Gaussian random variable \cite{berthier2013non, samanta2016chain,vandebroek2017effect,vandebroek2015dynamics,ghosh2014dynamics,
marconi2017heat}.

Initially, the system is in equilibrium with the thermal bath so the initial position $x_0$ is chosen from the Boltzmann distribution 
\begin{equation}
P(x_0,0)=\sqrt{\frac{k}{2\pi k_BT}} \exp\left(-\frac{\frac{1}{2}kx_0^2}{k_BT}\right)
\end{equation}
Then we can write $\left<x_0\right>=0$ and $\frac{1}{2}k\left<x_0^2\right>=\frac{1}{2}k_B T$, where $T$ is the temperature of the bath .

Using Laplace's transformation, we get the solution of Eq.(\ref{eq:langevinact}) for $t>0$ 
\begin{equation}
x(t)=x_0e^{-\frac{k}{\gamma}t}+\frac{1}{\gamma}\int_0^t dt^\prime e^{-\frac{k}{\gamma} (t-t^\prime)}\left(\xi (t^\prime) +\eta_A (t^\prime)\right)
\end{equation} 

At any time $t$ $(t > 0)$, the probability distribution is 
\begin{equation}
P(x,t)=\sqrt{\frac{1}{2\pi \left<x^2\right>}} \exp\left(-\frac{x^2}{2\left<x^2\right>}\right)
\end{equation}
where $\left<x^2\right>$ can be written as 
\begin{equation}
\left<x^2\right>=\frac{k_BT}{k}+\frac{k_BT_{act}}{k} \left(1-e^{-2\frac{k}{\gamma}t}\right)-\frac{2C}{\gamma^2(\frac{k^2}{\gamma^2}-\frac{1}{\tau_A^2})}\left(e^{-(\frac{k}{\gamma}+\frac{1}{\tau_A})t}-e^{-2\frac{k}{\gamma}t}\right)
\end{equation} 
\begin{equation*}
\begin{split}
\textrm{In the long time limit, one gets}\,\,\,\, \lim_{t \to \infty}\frac{1}{2}k\left<x^2\right> &=\frac{1}{2}k_B(T+T_{act})\,\,\,\textrm{where}\,\,\,T_{act}=\frac{C}{k_B\gamma {(\frac{k}{\gamma}+\frac{1}{\tau_A})}}\\
\end{split}
\end{equation*}
This is the generalized energy equipartition theorem in active bath  \cite{maggi2014generalized,samanta2016chain}. From this we can say, at long times, the system reaches a non-equilibrium steady state with an effective temperature $T_{eff}=(T+T_{act})$ \cite{samanta2016chain, vandebroek2015dynamics} as observed in earlier studies for Brownian particles \cite{wulfert2017driven,dieterich2015single} , where $T_{act}=\frac{C}{k_B\gamma {(\frac{k}{\gamma}+\frac{1}{\tau_A})}}$ is the effective temperature as obtained by Szamel for a harmonically trapped, self-driven, athermal particle \cite{szamel2014self}. But in our case we have a thermal system to begin with and then it is driven away from equilibrium. The readers are refereed to Appendix 1 for a detailed calculations.  
\subsection{Integral fluctuation theorem}
In the framework of stochastic thermodynamics, the first law is $Q = W_J-\Delta U$, where $Q$ is the heat being dissipated to the bath, $\Delta U$ is the change in internal energy and $W_J$ is the Jarzynski's work depends on external time dependent protocol \cite{sekimoto1998langevin}. In the same context, Seifert generalized the concept of entropy as the total entropy production along a single trajectory for a system driven out of equilibrium by time-dependent forces obeys the integral fluctuation theorem (IFT), $\left<e^{-\frac{\Delta S_{tot}}{k_B}}\right>=1$. In our case, $\Delta U=\frac{1}{2}kx^2 - \frac{1}{2}kx_0^2$ and $W_J=0$ for no external force. The change of entropy in the medium is $\Delta S_m = -\frac{\Delta U}{T}$ where $T$ is the unique ambient temperature of the medium, this definition holds even in the presence of active noise \cite{saha2009entropy}.

\noindent The change of entropy in the medium over the time interval $t$
\begin{equation}
\Delta S_m = -\left[\frac{\frac{1}{2}kx^2-\frac{1}{2}kx_0^2}{T}\right]
\end{equation}

\noindent The non-equilibrium entropy $(S(t))$ of the system is \cite{saha2009entropy, chaudhury2017}
\begin{equation}
S(t)=-k_B \int dx P(x,t) \ln P(x,t)=\left<s(t)\right>
\end{equation}
\noindent where $s(t)$ is the trajectory-dependent entropy of the system 
\begin{equation}
s(t)=-k_B \ln P(x(t),t)
\end{equation}
\noindent So the change in the entropy of the system for a trajectory during time $t$ 
\begin{equation}
\begin{split}
\Delta s &=-k_B \ln \left[ \frac{P(x,t)}{P(x_0,0)}\right]\\
&=-k_B \ln \left[\frac{\sqrt{\frac{1}{2\pi \left<x^2\right>}} \exp\left(-\frac{x^2}{2\left<x^2\right>}\right)}{\sqrt{\frac{k}{2\pi k_BT}} \exp\left(-\frac{kx_0^2}{2k_BT}\right)}\right]\\
\end{split}
\end{equation}
\begin{equation}
\begin{split}
\textrm{Total change in entropy}\,\,\,\Delta S_{tot}&=\Delta S_m + \Delta s  \\
&=-\frac{k_B}{2}\ln \left(\frac{k_BT}{k\left<x^2\right>}\right)+\left(\frac{k_B}{\left<x^2\right>}-\frac{k}{T}\right)\frac{1}{2}x^2 \\
&=a-\frac{bx^2}{2}
\end{split}
\end{equation}
\noindent where $a=-\frac{k_B}{2}\ln \left(\frac{k_BT}{k\left<x^2\right>}\right)$ and $b=\left(\frac{k}{T}-\frac{k_B}{\left<x^2\right>}\right)$
\\
So the total entropy production is quadratic function of $x$ and hence $\Delta S_{tot}$ is not Gaussian.
\\
One can write $P(\Delta S_{tot},t)=\int_{-\infty}^\infty dx P(x,t) \delta\left[\Delta S_{tot}-\left(a-\frac{bx^2}{2}\right)\right]$
\\
\noindent Analytical expression for the exact distribution of $P(\Delta S_{tot},t)$ is difficult. But the characteristic function of $P(\Delta S_{tot},t)$ can easily be found. The characteristic form of the $P(\Delta S_{tot},t)$ is defined as
\begin{equation}
\tilde{P}(R,t) \equiv \int_{-\infty}^\infty d\Delta S_{tot} e^{iR\Delta S_{tot}} P(\Delta S_{tot},t)
\end{equation} 
\noindent From that we can get the IFT as $\left<e^{\frac{-\Delta S_{tot}}{k_B}}\right>=\tilde{P}(i/k_B,t)=1$ \cite{van2004extended, saha2009entropy}.
\\
\noindent Thus we get $\tilde{P}(R,t)=e^{iRa}\sqrt{\frac{1}{\left(1+iRb\left<x^2\right>\right)}}$

\noindent For a detailed calculation see Appendix 2.
\begin{equation}
\begin{split}
\textrm{So}\,\,\,\,\tilde{P}(i/k_B,t)&=e^{-\frac{a}{k_B}}\sqrt{\frac{1}{\left(1-\frac{b\left<x^2(t)\right>}{k_B}\right)}}\\
&=\left[{\frac{k_BT}{k\left<x^2(t)\right>}}\right] \times {\left[{\frac{2k_BT}{k\left<x^2(t)\right>}-1}\right]}^{-\frac{1}{2}}\\
& \neq 1
\label{eq:IFT}
\end{split}
\end{equation}
\noindent Thus, IFT cannot be applied in our case. But at $t=0$, the system is in thermal equilibrium with the medium and $x=x_0$. Using the equipartition theorem at thermal equilibrium $\left<x_0^2\right>=\frac{k_BT}{k}$ in Eq.(\ref{eq:IFT}), we get $\tilde{P}(-i/k_B,t)=1$ and IFT is recovered. 
 
\section{Work distribution for a colloidal particle in an active bath subjected to a constant drag}
Here we consider a protocol in which the center of the harmonic well is moved with a constant velocity $u$. In this case the Hamiltonian of the system is time dependent
\begin{equation}
H(t)=\frac{p^2}{2m}+\frac{1}{2}k(x-f(t))^2
\end{equation}
Where $f(t)=ut$ is the external time-dependent protocol and the dynamics of the particle is governed by the overdamped Langevin equation 
\begin{equation}
\gamma \dot x = -k(x-f(t)) + \xi(t) + \eta_A(t)
\label{eq:langevinexternal}
\end{equation}
This type of model has been studied theoretically \cite{dhar2005work} as well as has an experimental relevance \cite{argun2016non, maggi2014generalized, trepagnier2004experimental}.

Initially at time $t=0$, the system is in equilibrium with the thermal bath so the initial position $x_0$ is chosen from the Boltzmann distribution $P(x_0) \sim e^{-\beta H_0}$ where $H_0=\frac{1}{2}kx_0^2$, $\beta=\frac{1}{k_B T}$. This ensures $\left<x_0\right>=0$ and $\frac{1}{2}k\left<x_0^2\right>=\frac{1}{2}k_B T$ , is the equipartition theorem. 

Using Laplace's transformation, we get the solution of Eq.(\ref{eq:langevinexternal}) for $t>0$ 
\begin{equation}
\begin{split}
x(t)&=x_0e^{-\frac{k}{\gamma}t}+\frac{1}{\gamma}\int_0^t dt^\prime e^{-\frac{k}{\gamma} (t-t^\prime)}\left(kf(t^\prime) + \xi (t^\prime) +\eta_A (t^\prime)\right)\\
&=x_0e^{-\frac{k}{\gamma}t}+\frac{1}{\gamma}\int_0^t dt^\prime e^{-\frac{k}{\gamma} (t-t^\prime)}\left(kut^\prime + \xi (t^\prime) +\eta_A (t^\prime)\right)
\label{eq:langevinexternalsoln}
\end{split}
\end{equation} 
The average position $\left<x(t)\right>=\frac{ku}{\gamma}\int_0^t dt^\prime e^{-\frac{k}{\gamma} (t-t^\prime)}t^\prime$
\subsection{Jarzynski's work}
Following the definition of Jarzynski's work done as used in the context of stochastic thermodynamics subjected to the above mentioned protocol over a finite duration $t$ \cite{jarzynski1997nonequilibrium} we can write
\begin{equation}
\begin{split}
W_J&=\int_0^t \frac{\partial H}{\partial f(t^\prime)} \dot{f}(t^\prime) dt^\prime\\
&=k\int_0^t dt^\prime \dot{f}(t^\prime) f(t^\prime)-k\int_0^t dt^\prime \dot{f}(t^\prime)x(t^\prime)
\label{eq:Jarzynskiwork}
\end{split}
\end{equation}
In our case this is equal to the mechanical work $(W_m=\int_0^t dt^\prime f(t^\prime) \dot{x}(t^\prime))$ done on the particle \cite{narayan2003reexamination}.
Thus, $W_J=W_m=W$ is the work done on the particle. The full calculation can be found in Appendix 3.

Here $x_0$, $\xi$ and $\eta_A$ are Gaussian random variables. The work $(W)$ is a linear functional of $x$ and $x$ is a linear combination stochastic variables are $x_0$, $\xi$ and $\eta_A$. Therefore the work distribution function is Gaussian and the mean and variance is sufficient to find the exact distribution. 
\subsection{Transient work fluctuation theorem}}
The conventional form of transient work fluctuation relation is $\frac{P(W)}{P(-W)}=e^{A W}$ where $A$ is a constant which depends neither on $t$, nor on $W$ \cite{touchette2007fluctuation}. In case of an unique ambient medium temperature $T$, $A$ reduces to conventional  $\beta\left(=\frac{1}{k_B T}\right)$ \cite{chakrabarti2009transient,mai2007nonequilibrium}. The distribution functions for the positive and the negative work respectively are $P(W)=\frac{1}{\sqrt{2\pi \sigma_W^2}}\exp\left[-\frac{(W-\left<W\right>)^2}{2\sigma_W^2}\right]$ and $P(-W)=\frac{1}{\sqrt{2\pi \sigma_W^2}}\exp\left[-\frac{(W+\left<W\right>)^2}{2\sigma_W^2}\right]$, which immediately leads to $\frac{P(W)}{P(-W)}=\exp\left[{\frac{2\left<W\right>W}{\sigma_W^2}}\right]$. For Gaussian work distributions, the condition for variance, $\sigma_W^2=2 k_B T \left<W\right> $, guarantees transient fluctuation theorem in conventional form, $\frac{P(W)}{P(-W)}=e^{\frac{W}{k_B T}}$ \cite{ritort2008nonequilibrium}

The average work for the time duration $t$,
\begin{equation}
\left<W\right>=u^2 \gamma \left[t +\frac{\gamma}{k}\left(e^{-\frac{k}{\gamma}t}-1\right)\right]
\end{equation}
For a detailed calculation  see Appendix 4.
\begin{equation}
\begin{split}
\textrm{The variance for work}\,\,\,\sigma_W^2 &=\left<(W-\left<W\right>)^2\right>\\
&=k^2u^2\int_{0}^{t} dt_1\int_{0}^{t} dt_2 \left<\Delta x(t_1) \Delta x(t_2)\right>\\
&=k^2u^2\int_{0}^{t} dt_1\int_{0}^{t} dt_2 h(t_1,t_2)
\label{eq:w(t)}
\end{split}
\end{equation}
here $\Delta x(t_1)=x(t_1)-\left<x(t_1)\right>$ and $h(t_1,t_2)=\left<\Delta x(t_1) \Delta x(t_2)\right>$
\\
\begin{equation}
\begin{split}
\textrm{where}\,\,\,\,h(t_1,t_2)&=e^{-\frac{k}{\gamma} (t_1+t_2)}\left<x_0^2\right>+\frac{1}{\gamma^2}\int_0^{t_1} dt^\prime e^{-\frac{k}{\gamma} (t_1-t^\prime)}\int_0^{t_2} dt^{\prime\prime} e^{-\frac{k}{\gamma} (t_2-t^{\prime\prime})}\left<\xi (t^\prime)\xi (t^{\prime\prime})\right>\\
&+\frac{1}{\gamma^2}\int_0^{t_1} dt^\prime e^{-\frac{k}{\gamma} (t_1-t^\prime)}\int_0^{t_2} dt^{\prime\prime} e^{-\frac{k}{\gamma} (t_2-t^{\prime\prime})}\left<\eta_A (t^\prime)\eta_A (t^{\prime\prime})\right>\\
&=e^{-\frac{k}{\gamma} (t_1+t_2)}\left<x_0^2\right>+\frac{2k_BT\gamma}{\gamma^2}\int_0^{t_1} dt^\prime e^{-\frac{k}{\gamma} (t_1-t^\prime)}\int_0^{t_2} dt^{\prime\prime} e^{-\frac{k}{\gamma} (t_2-t^{\prime\prime})}\delta(t^\prime-t^{\prime\prime})\\
&+\frac{C}{\gamma^2}\int_0^{t_1} dt^\prime e^{-\frac{k}{\gamma} (t_1-t^\prime)}\int_0^{t_2} dt^{\prime\prime} e^{-\frac{k}{\gamma} (t_2-t^{\prime\prime})}e^{-\frac{|t^\prime-t^{\prime\prime}|}{\tau_A}}\\
&=\frac{k_BT}{k}e^{-\frac{k}{\gamma} (t_1-t_2)}+\frac{C}{\gamma^2}\Bigg[\frac{e^{-\frac{1}{\tau_A}(t_1-t_2)}-e^{-\frac{k}{\gamma}(t_1-t_2)}-e^{-(\frac{k}{\gamma}t_2+\frac{1}{\tau_A}t_1)}-e^{-(\frac{k}{\gamma}t_1+\frac{1}{\tau_A}t_2)}}{(\frac{k}{\gamma}-\frac{1}{\tau_A})(\frac{k}{\gamma}+\frac{1}{\tau_A})}\\
&+\frac{e^{-\frac{k}{\gamma}(t_1-t_2)}}{\frac{k}{\gamma}(\frac{k}{\gamma}+\frac{1}{\tau_A})}+\frac{e^{-\frac{k}{\gamma} (t_1+t_2)}}{\frac{k}{\gamma}\left(\frac{k}{\gamma}-\frac{1}{\tau_A}\right)}\Bigg]
\label{eq:h(t)}
\end{split}
\end{equation}
For a detailed calculation see Appendix 5.

\noindent Substituing Eq.(\ref{eq:h(t)}) in Eq.(\ref{eq:w(t)}) we get
\begin{equation}
\begin{split}
\sigma_W^2 &=k^2u^2\int_{0}^{t} dt_1\int_{0}^{t} dt_2 h(t_1,t_2)\\
&=2k_BT_{eff}\left<W\right>+\frac{2k^2u^2C}{\gamma^2 (\frac{k}{\gamma}-\frac{1}{\tau_A})(\frac{k}{\gamma}+\frac{1}{\tau_A})}\Bigg[\frac{\gamma\tau_A}{k}\left(\frac{k}{\gamma}-\frac{1}{\tau_A}\right)t+\left(\tau_A^2e^{-\frac{1}{\tau_A}t}-\frac{\gamma^2}{k^2}e^{-\frac{k}{\gamma} t}\right)\\
&-\left(\tau_A^2-\frac{\gamma^2}{k^2}\right)-\frac{\gamma\tau_A}{k}e^{-(\frac{k}{\gamma}+\frac{1}{\tau_A})t}+\frac{\gamma\tau_A}{k}\left(e^{-\frac{1}{\tau_A}t}+e^{-\frac{k}{\gamma}t}\right)-\frac{\gamma\tau_A}{k}\Bigg]\\
&+\frac{\gamma}{k}\frac{u^2C} {(\frac{k}{\gamma}-\frac{1}{\tau_A})}\left(e^{-2\frac{k}{\gamma} t}-1\right)-\frac{\gamma}{k}\frac{2u^2C} {(\frac{k}{\gamma}-\frac{1}{\tau_A})}\left(e^{-\frac{k}{\gamma} t}-1\right)\\
&=2k_BT_{eff}\left<W\right>+Z(t)
\end{split}
\end{equation}
\begin{equation}
\begin{split}
\textrm{Where}\,\,\,\,\ Z(t)&=\frac{2k^2u^2C}{\gamma^2 (\frac{k}{\gamma}-\frac{1}{\tau_A})(\frac{k}{\gamma}+\frac{1}{\tau_A})}\Bigg[\frac{\gamma\tau_A}{k}\left(\frac{k}{\gamma}-\frac{1}{\tau_A}\right)t+\left(\tau_A^2e^{-\frac{1}{\tau_A}t}-\frac{\gamma^2}{k^2}e^{-\frac{k}{\gamma} t}\right)\\
&-\left(\tau_A^2-\frac{\gamma^2}{k^2}\right)-\frac{\gamma\tau_A}{k}e^{-(\frac{k}{\gamma}+\frac{1}{\tau_A})t}+\frac{\gamma\tau_A}{k}\left(e^{-\frac{1}{\tau_A}t}+e^{-\frac{k}{\gamma}t}\right)-\frac{\gamma\tau_A}{k}\Bigg]\\
&+\frac{\gamma}{k}\frac{u^2C} {(\frac{k}{\gamma}-\frac{1}{\tau_A})}\left(e^{-2\frac{k}{\gamma} t}-1\right)-\frac{\gamma}{k}\frac{2u^2C} {(\frac{k}{\gamma}-\frac{1}{\tau_A})}\left(e^{-\frac{k}{\gamma} t}-1\right)\\
\end{split}
\end{equation}
For a detailed calculation see Appendix 6.
 
\noindent This results
 \begin{equation}
\begin{split}
\frac{P(W)}{P(-W)}&=e^{\frac{2\left<W\right>W}{\sigma_W^2}}\\
&=e^{\frac{2\left<W\right>W}{2k_BT_{eff}\left<W\right>+Z(t)}}\\
&=e^{I(t)W}\\
\end{split}
\end{equation}

where $I(t)=\frac{2\left<W\right>}{2k_BT_{eff}\left<W\right>+Z(t)}$
\\
\noindent So the conventional TFT for work cannot be applied in this case. This is again a consequence of the fact that the initial states for the forward and the backward processes are sampled from different distributions. 
\subsection{Steady state work fluctuation theorem} 
To investigate the SSFT, we consider an arbitrary time $t_0$ when the system is in an arbitrary initial condition and then drive the system to a steady state \cite{ciliberto2010fluctuations}. 

Using Eq.(\ref{eq:langevinexternalsoln}) and Eq.(\ref{eq:Jarzynskiwork}) the average work for a time duration $t$,
\begin{equation}
\begin{split}
\left<W\right>&=k\int_{t_0}^{t+t_0}dt^\prime\frac{d}{dt^\prime}\left(\frac{f^2(t^\prime)}{2}\right) -ku\int_{t_0}^{t+t_0}dt^\prime\left<x(t^\prime)\right>\\
&=u^2 \gamma t+\frac{\gamma^2u^2}{k}\left(e^{-\frac{k}{\gamma}(t+t_0)}-e^{-\frac{k}{\gamma}t_0}\right)
\end{split}
\end{equation}
The initial time $t_0$ is sufficiently large so the particle is perpetually in the steady state
\begin{equation}
\lim_{t_0 \to \infty}\left<W\right>=u^2 \gamma t
\end{equation}

Similar to the section III.B, the variance for work is given by  
\begin{equation}
\begin{split}
\sigma_{W_{ss}}^2 &=\left<(W-\left<W\right>)^2\right>\\
&=k^2u^2\int_{t_0}^{t+t_0} dt_1\int_{t_0}^{t+t_0} dt_2 h(t_1,t_2)\\
&=k^2u^2\int_{0}^{t} dt_1\int_{0}^{t} dt_2 h(t_1+t_0,t_2+t_0)
\end{split}
\end{equation}

\begin{equation}
\begin{split}
\textrm{Now}\,\,\,\,h(t_1,t_2)&=e^{-\frac{k}{\gamma} (t_1+t_2)}\left(\left<x_0^2\right>-\frac{k_BT}{k}\right)+\frac{k_BT}{k}e^{-\frac{k}{\gamma} (t_1-t_2)}\\
&+\frac{C}{\gamma^2}e^{-\frac{k}{\gamma} (t_1+t_2)}\int_0^{t_1} dt^\prime\int_0^{t_2} dt^{\prime\prime} e^{\frac{k}{\gamma} (t^\prime+t^{\prime\prime})}e^{-\frac{|t^\prime-t^{\prime\prime}|}{\tau_A}}\\
&=e^{-\frac{k}{\gamma} (t_1+t_2)}\left(\left<x_0^2\right>-\frac{k_BT}{k}\right)+\frac{k_BT}{k}e^{-\frac{k}{\gamma} (t_1-t_2)}\\
&+\frac{C}{\gamma^2}\Bigg[\frac{e^{-\frac{1}{\tau_A}(t_1-t_2)}-e^{-\frac{k}{\gamma}(t_1-t_2)}-e^{-(\frac{k}{\gamma}t_2+\frac{1}{\tau_A}t_1)}-e^{-(\frac{k}{\gamma}t_1+\frac{1}{\tau_A}t_2)}}{(\frac{k}{\gamma}-\frac{1}{\tau_A})(\frac{k}{\gamma}+\frac{1}{\tau_A})}\\
&+\frac{e^{-\frac{k}{\gamma}(t_1-t_2)}}{\frac{k}{\gamma}(\frac{k}{\gamma}+\frac{1}{\tau_A})}+\frac{e^{-\frac{k}{\gamma} (t_1+t_2)}}{\frac{k}{\gamma}\left(\frac{k}{\gamma}-\frac{1}{\tau_A}\right)}\Bigg]
\end{split}
\end{equation}

\begin{equation}
\begin{split}
\textrm{Again}\,\,\,\,  h(t_1+t_0,t_2+t_0)&=e^{-\frac{k}{\gamma} (t_1+t_0+t_2+t_0)}\left(\left<x_0^2\right>-\frac{k_BT}{k}\right)+\frac{k_BT}{k}e^{-\frac{k}{\gamma} (t_1-t_2)}\\
&+{C}{\gamma^2}\Bigg[\frac{e^{-\frac{1}{\tau_A}(t_1-t_2)}-e^{-\frac{k}{\gamma}(t_1-t_2)}-e^{-(\frac{k}{\gamma}(t_2+t_0)+\frac{1}{\tau_A}(t_1+t_0))}-e^{-(\frac{k}{\gamma}(t_1+t_0)+\frac{1}{\tau_A}(t_2+t_0))}}{(\frac{k}{\gamma}-\frac{1}{\tau_A})(\frac{k}{\gamma}+\frac{1}{\tau_A})}\\
&+\frac{e^{-\frac{k}{\gamma}(t_1-t_2)}}{\frac{k}{\gamma}(\frac{k}{\gamma}+\frac{1}{\tau_A})}+\frac{e^{-\frac{k}{\gamma} (t_1+t_0+t_2+t_0)}}{\frac{k}{\gamma}\left(\frac{k}{\gamma}-\frac{1}{\tau_A}\right)}\Bigg]
\end{split}
\end{equation}

\noindent In the limit $t_0 \rightarrow \infty$ 
\begin{equation}
\begin{split}
h(t_1+t_0,t_2+t_0)&=\frac{k_BT}{k}e^{-\frac{k}{\gamma} (t_1-t_2)}+\frac{C}{\gamma^2}\left[\frac{e^{-\frac{1}{\tau_A}(t_1-t_2)}-e^{-\frac{k}{\gamma}(t_1-t_2)}}{(\frac{k}{\gamma}-\frac{1}{\tau_A})(\frac{k}{\gamma}+\frac{1}{\tau_A})}+\frac{e^{-\frac{k}{\gamma}(t_1-t_2)}}{\frac{k}{\gamma}(\frac{k}{\gamma}+\frac{1}{\tau_A})}\right]\\
&=\frac{k_BT}{k}e^{-\frac{k}{\gamma} (t_1-t_2)}+\frac{C}{\gamma{(\frac{k}{\gamma}+\frac{1}{\tau_A})} {(\frac{k}{\gamma}-\frac{1}{\tau_A})}} \left[\frac{e^{-\frac{1}{\tau_A}(t_1-t_2)}}{\gamma} -\frac{e^{-\frac{k}{\gamma}(t_1-t_2)}}{k\tau_A}\right]
\end{split}
\end{equation} 

\begin{equation}
\begin{split}
\textrm{Now,}\,\,\,\,\sigma_{W_{ss}}^2 &=k^2u^2\int_{0}^{t} dt_1\int_{0}^{t} dt_2 h(t_1+t_0,t_2+t_0)\\
&=k^2u^2\int_{0}^{t} dt_1\int_{0}^{t} dt_2\left[\frac{k_BT}{k}e^{-\frac{k}{\gamma} (t_1-t_2)}+\frac{C}{\gamma{(\frac{k}{\gamma}+\frac{1}{\tau_A})} {(\frac{k}{\gamma}-\frac{1}{\tau_A})}} \left(\frac{e^{-\frac{1}{\tau_A}(t_1-t_2)}}{\gamma} -\frac{e^{-\frac{k}{\gamma}(t_1-t_2)}}{k\tau_A}\right)\right]\\
&=2k_BT u^2\gamma t \left[1-\frac{\gamma}{kt}\left(e^{-\frac{k}{\gamma}t}-1\right)\right]+\frac{2k^2u^2 \tau_A Ct}{ {\gamma^2{(\frac{k}{\gamma}+\frac{1}{\tau_A})}(\frac{k}{\gamma}-\frac{1}{\tau_A})}}\left[1-\frac{1}{t\tau_A} \left(e^{-\frac{1}{\tau_A}t}-1\right)\right] \\
&-\frac{2u^2 \gamma Ct}{ {\gamma\tau_A{(\frac{k}{\gamma}+\frac{1}{\tau_A})}(\frac{k}{\gamma}-\frac{1}{\tau_A})}} \left[1-\frac{\gamma}{kt}\left(e^{-\frac{k}{\gamma}t}-1\right)\right] \Bigg] 
\end{split}
\end{equation}
\noindent For a detailed calculation  see Appendix 7.

\noindent In the long time limit $t \rightarrow \infty$, 
\begin{equation}
\begin{split}
\sigma_{W_{ss}}^2 &=2k_BT u^2\gamma t+\frac{2k^2u^2 \tau_A Ct}{ {\gamma^2{(\frac{k}{\gamma}+\frac{1}{\tau_A})}(\frac{k}{\gamma}-\frac{1}{\tau_A})}}-\frac{2u^2 \gamma Ct}{ {\gamma\tau_A{(\frac{k}{\gamma}+\frac{1}{\tau_A})}(\frac{k}{\gamma}-\frac{1}{\tau_A})}}\\
&=2\left[k_BT+\frac{C \tau_A}{\gamma} \right]u^2 \gamma t\\
&=2\left[k_BT+\frac{C \tau_A}{\gamma} \right]\left<W\right>
\end{split}
\end{equation} 
\begin{equation}
\begin{split}
\textrm{Let}\,\, \omega=\frac{W}{t},\,\,\,\textrm{Then}\,\,\,\,P_s (\omega)&=tP(W)\\
&=\frac{t}{\sqrt{2\pi \sigma_W^2}}\exp\left[-\frac{(\omega-\left<\omega\right>)^2 t}{2  \frac{\sigma_W^2}{t}}\right]
\end{split}
\end{equation} 
\noindent This means $\frac{P(\omega)}{P(-\omega)}=e^{2\frac{\left<\omega\right>}{\sigma_W^2/t} \omega t}$ and in the limit $t \rightarrow \infty$, one gets
\begin{equation}
\begin{split}
\lim_{t \rightarrow \infty} \frac{1}{t} \ln \frac{P(\omega)}{P(-\omega)}&=2\frac{\left<\omega\right>}{\sigma_W^2/t} \omega \\
&=2\frac{\left<W\right>/t}{\sigma_W^2/t} \omega \\
&=\frac{\omega}{k_B\left[T+\frac{C \tau_A}{k_B\gamma} \right]}  \\
&=\omega \alpha  
\end{split}
\end{equation} 
\noindent Where $\alpha =\frac{1}{k_B\left[T+\frac{C \tau_A}{k_B\gamma} \right]} $. Notice that $\alpha$ is nothing but the effective inverse temperature  with no trapping $(k=0)$ \cite{wu2000particle}. We believe that a properly designed experiment should be able to verify this in future. We also like to point out that Sen $et\,\,al$ had arrived at a similar result but in a different context \cite{sen2011work}.  In absence of active noise $(C=0)$, $\alpha$ reduced to $\beta=\frac{1}{k_B T} $  and we recover the steady state fluctuation theorem of Gallavoti and Cohen \cite{gallavotti1995dynamical}.

\section{Conclusions}
Motivated by recent claims \cite{argun2016non} that fluctuation relations cannot be directly applied in case of a passive colloid in an active bath, here we consider a model system of a harmonically trapped particle subjected to a thermal and an active noise. In our model, the active forces are characterized by Gaussian random variable with zero mean and an exponentially decaying temporal correlation. At long times, the system reaches a non-equilibrium steady state which can be described using a Boltzmann type distribution but with an effective temperature different from the ambient temperature. We obtain analytical expressions of the probability density function for work subjected to constant dragging (Gaussian) and entropy in the absence of dragging (non-Gaussian). Our analysis shows that if the system is initially at thermal equilibrium, then by adding active particles in the medium, entropy is continuously produced. In this case, the active force cannot be connected to the friction because of the absence of any fluctuation-dissipation theorem.
\noindent The key finding of our work is that IFT for entropy and  TFT for work  cannot be applied for this system similar to the case of Levy noise \cite{touchette2007fluctuation}. However SSFT for work exists with $\beta\left(=\frac{1}{k_B T}\right)$ being renormalized to $\alpha \left(=\left[k_B\left(T+\frac{C \tau_A}{k_B\gamma} \right)\right]^{-1} \right)$. This is because of  generalized energy equipartition theorem for active systems in steady state \cite{maggi2014generalized,sen2011work}. 
\\
We would like to point out that, if the trap relaxation time $(\frac{\gamma}{k})$ is greater than the bacterial correlation time $(\tau_A)$ such as in case of weak trapping or high viscosity, we can apply the Gaussian approximation to the active forces as the higher moments are not important in this case. But if the two time scales are comparable or the bacterial correlation time is higher than the trap relaxation time, then the active forces are no longer Gaussian. One can still define the effective temperature for active systems with non-Gaussian distribution as $T_{eff}=\frac{k\left<x^2\right>}{k_B}$ where $\left<x^2\right>$ is the variance of that non-Gaussian distribution \cite{krishnamurthy2016micrometre}. A further challenge would be to come up with the notion of effective temperature for a energy landscape with multiple minima, each of which corresponds to a steady state (metastable). In a future work, it will be interesting to investigate this regime of non-Gaussian fluctuations.

\section{Acknowledgments} 
We would like to acknowledge Prof. Abhishek Dhar and Prof. K. L. Sebastian for useful comments and discussions. RC acknowledges SERB for financial support (Project No. SB/SI/PC-55/2013). SC acknowledges DST-Inspire for the fellowship.

\section{Appendix}
\subsubsection{Calculation of MSD} 
\begin{equation}
\begin{split}
\left<x^2(t)\right>&=\left<x_0^2\right>e^{-2\frac{k}{\gamma}t}+\frac{1}{\gamma^2}\int_0^{t} dt^\prime e^{-\frac{k}{\gamma} (t-t^\prime)}\int_0^{t} dt^{\prime\prime} e^{-\frac{k}{\gamma} (t-t^{\prime\prime})}\left[\left<\xi (t^\prime)\xi (t^{\prime\prime})\right>+\left<\eta_A (t^\prime)\eta_A (t^{\prime\prime})\right>\right]\\
&=\frac{k_BT}{k} e^{-2\frac{k}{\gamma}t}+\frac{2k_BT\gamma e^{-2\frac{k}{\gamma}t}}{\gamma^2}\int_0^{t} dt^\prime \int_0^{t} dt^{\prime\prime} e^{\frac{k}{\gamma} (t^\prime+t^{\prime\prime})}\delta(t^\prime-t^{\prime\prime}) \\
&+\frac{Ce^{-2\frac{k}{\gamma}t}}{\gamma^2}\int_0^{t} dt^\prime \int_0^{t} dt^{\prime\prime} e^{\frac{k}{\gamma} (t^\prime+t^{\prime\prime})} e^{-\frac{|t^\prime-t^{\prime\prime}|}{\tau_A}}\\
&=\frac{k_BT}{k}+\frac{C}{k\gamma {(\frac{k}{\gamma}+\frac{1}{\tau_A})}} \left(1-e^{-2\frac{k}{\gamma}t}\right)-\frac{2C}{\gamma^2(\frac{k^2}{\gamma^2}-\frac{1}{\tau_A^2})}\left(e^{-(\frac{k}{\gamma}+\frac{1}{\tau_A})t}-e^{-2\frac{k}{\gamma}t}\right)\\
&=\frac{k_BT}{k}+\frac{k_BT_{act}}{k} \left(1-e^{-2\frac{k}{\gamma}t}\right)-\frac{2C}{\gamma^2(\frac{k^2}{\gamma^2}-\frac{1}{\tau_A^2})}\left(e^{-(\frac{k}{\gamma}+\frac{1}{\tau_A})t}-e^{-2\frac{k}{\gamma}t}\right)
\end{split}
\end{equation}
Where $T_{act}=\frac{C}{k_B\gamma {(\frac{k}{\gamma}+\frac{1}{\tau_A})}}$

\noindent Then $\lim_{t \to \infty}\left<x^2(t)\right> =\frac{k_BT}{k}+\frac{k_BT_{act}}{k}=\frac{k_B T_{eff}}{k}$

\subsubsection{Calculation of characteristic function of $P(\Delta S_{tot},t)$}
\begin{equation}
\begin{split}
\tilde{P}(R,t) &\equiv \int_{-\infty}^\infty d\Delta S_{tot} e^{iR\Delta S_{tot}} P(\Delta S_{tot},t)\\
&=\int_{-\infty}^\infty d\Delta S_{tot} e^{iR\Delta S_{tot}} \int_{-\infty}^\infty dx P(x,t) \delta \left[\Delta S_{tot}-(a-\frac{bx^2}{2})\right]\\
&=\int_{-\infty}^\infty dx P(x,t)\int_{-\infty}^\infty d\Delta S_{tot} e^{iR\Delta S_{tot}} \delta \left[\Delta S_{tot}-(a-\frac{bx^2}{2})\right] \\
&=\int_{-\infty}^\infty dx P(x,t) e^{iR(a-\frac{bx^2}{2})} \\
&=e^{iRa}\int_{-\infty}^\infty dx P(x,t) e^{-\frac{iRbx^2}{2}}\\
&=e^{iRa}\sqrt{\frac{1}{2\pi \left<x^2\right>}}\int_{-\infty}^\infty dx \exp\left[-\frac{1}{2}\left(\frac{1}{\left<x^2\right>}+iRb\right)x^2\right]\\
&=e^{iRa}\sqrt{\frac{1}{2\pi \left<x^2\right>}} \sqrt{\frac{2\pi}{\left(\frac{1}{\left<x^2\right>}+iRb\right)}} \,\,\,,b>0\\
&=e^{iRa}\sqrt{\frac{1}{\left(1+iRb\left<x^2\right>\right)}}
\end{split}
\end{equation}

\subsubsection{Calculation of Jarzynski's work and mechanical work}
For a dragged harmonic oscillator, the Jarzynski's work is given by 
\begin{equation}
\begin{split}
W_J&=\int_0^t \frac{\partial H}{\partial f(t^\prime)} \dot{f}(t^\prime) dt^\prime\\
&=-k \int_0^t \dot{f}(t^\prime)  (x-f(t^\prime))dt^\prime\\
&=k\int_0^t dt^\prime \dot{f}(t^\prime) f(t^\prime)-k\int_0^t dt^\prime \dot{f}(t^\prime)x(t^\prime)
\end{split}
\end{equation}
The mechanical work done on the particle is given by 
\begin{equation}
\begin{split}
W_m &=\int_0^t dt^\prime f(t^\prime) \dot{x}(t^\prime)\\
&=-\int_0^t dt^\prime \dot{x}(t^\prime)  \partial_x H \\
&=-\int_0^t dt^\prime \partial_t H \\
&=k\int_0^t dt^\prime \dot{f}(t^\prime) f(t^\prime)-k\int_0^t dt^\prime \dot{f}(t^\prime)x(t^\prime)
\end{split}
\end{equation}
So $W_J=W_m$.

\subsubsection{Calculation of average position and average work}
\begin{equation}
\begin{split}
\left<x(t)\right>&=\frac{1}{\gamma}\int_0^t dt^\prime e^{-\frac{k}{\gamma} (t-t^\prime)}kf(t^\prime)\\
&=\frac{ku}{\gamma}\int_0^t dt^\prime e^{-\frac{k}{\gamma} (t-t^\prime)} t^\prime\\
&=\frac{ku}{\gamma}e^{-\frac{k}{\gamma}t}\int_0^t dt^\prime e^{\frac{k}{\gamma} t^\prime} t^\prime\\ 
&=\frac{ku}{\gamma}e^{-\frac{k}{\gamma}t}\left[\frac{\gamma}{k}\left(t^\prime e^{-\frac{k}{\gamma}t^\prime}\right)_0^t-\frac{\gamma^2}{k^2}\left(e^{\frac{k}{\gamma}t^\prime}-1\right)_0^t\right]\\
&=ut-\frac{\gamma u}{k}+\frac{\gamma u}{k}e^{-\frac{k}{\gamma}t}
\end{split}
\end{equation}
\begin{equation}
\begin{split}
\left<W\right>&=k\int_0^t dt^\prime \dot{f}(t^\prime) f(t^\prime)-k\int_0^t dt^\prime \dot{f}(t^\prime)\left<x(t^\prime)\right>\\
&=k\int_0^t dt^\prime \frac{d}{dt^\prime}\left(\frac{f^2(t^\prime)}{2}\right)-ku\int_0^t dt^\prime \left<x(t^\prime)\right> \\
&=\frac{k}{2}\left[f^2(t)-f^2(0)\right]-ku\int_0^t dt^\prime \left[ut^\prime-\frac{\gamma u}{k}+\frac{\gamma u}{k}e^{-\frac{k}{\gamma}t^\prime} \right] \\
&=\frac{ku^2t^2}{2}-\frac{ku^2t^2}{2}+u^2\gamma t +\frac{\gamma^2u^2}{k}\left[e^{-\frac{k}{\gamma}t}-1\right] \\
&= u^2 \gamma \left[t +\frac{\gamma}{k}\left(e^{-\frac{k}{\gamma}t}-1\right)\right]
\end{split}
\end{equation}
\subsubsection{Calculation of the two point correlation function}
The two point correlation function of the position of Brownian particle, $h(t_1,t_2)$, is given by
\begin{equation}
\begin{split}
h(t_1,t_2)&=\left<(x(t_1)-\left<x(t_1)\right>)(x(t_2)-\left<x(t_2)\right>)\right>\\
&=\Bigg<\left[x_0e^{-\frac{k}{\gamma}t_1}+\frac{1}{\gamma}\int_0^{t_1} dt^\prime e^{-\frac{k}{\gamma} (t_1-t^\prime)}\left(\xi (t^\prime) +\eta_A (t^\prime)\right)\right]\\
&\times \left[x_0e^{-\frac{k}{\gamma}t_2}+\frac{1}{\gamma}\int_0^{t_2} dt^{\prime\prime} e^{-\frac{k}{\gamma} (t_2-t^{\prime\prime})}\left(\xi (t^{\prime\prime}) +\eta_A (t^{\prime\prime})\right)\right]\Bigg>\\
&=e^{-\frac{k}{\gamma} (t_1+t_2)}\left<x_0^2\right>+\frac{1}{\gamma^2}\int_0^{t_1} dt^\prime e^{-\frac{k}{\gamma} (t_1-t^\prime)}\int_0^{t_2} dt^{\prime\prime} e^{-\frac{k}{\gamma} (t_2-t^{\prime\prime})}\left<\xi (t^\prime)\xi (t^{\prime\prime})\right>\\
&+\frac{1}{\gamma^2}\int_0^{t_1} dt^\prime e^{-\frac{k}{\gamma} (t_1-t^\prime)}\int_0^{t_2} dt^{\prime\prime} e^{-\frac{k}{\gamma} (t_2-t^{\prime\prime})}\left<\eta_A (t^\prime)\eta_A (t^{\prime\prime})\right>\\
&=e^{-\frac{k}{\gamma} (t_1+t_2)}\left<x_0^2\right>+\frac{2k_BT\gamma}{\gamma^2}\int_0^{t_1} dt^\prime e^{-\frac{k}{\gamma} (t_1-t^\prime)}\int_0^{t_2} dt^{\prime\prime} e^{-\frac{k}{\gamma} (t_2-t^{\prime\prime})}\delta(t^\prime-t^{\prime\prime})\\
&+\frac{C}{\gamma^2}\int_0^{t_1} dt^\prime e^{-\frac{k}{\gamma} (t_1-t^\prime)}\int_0^{t_2} dt^{\prime\prime} e^{-\frac{k}{\gamma} (t_2-t^{\prime\prime})}e^{-\frac{|t^\prime-t^{\prime\prime}|}{\tau_A}}\\
&=\frac{k_BT}{k}e^{-\frac{k}{\gamma} (t_1-t_2)}+\frac{C}{\gamma^2}e^{-\frac{k}{\gamma} (t_1+t_2)}\int_0^{t_1} dt^\prime\int_0^{t_2} dt^{\prime\prime} e^{\frac{k}{\gamma} (t^\prime+t^{\prime\prime})}e^{-\frac{|t^\prime-t^{\prime\prime}|}{\tau_A}} \\
&=\frac{k_BT}{k}e^{-\frac{k}{\gamma} (t_1-t_2)}+\frac{C}{\gamma^2}e^{-\frac{k}{\gamma} (t_1+t_2)} \times K
\label{eq:h(t1t2)}
\end{split}
\end{equation}
Where $K=\int_0^{t_1} dt^\prime\int_0^{t_2} dt^{\prime\prime} e^{\frac{k}{\gamma} (t^\prime+t^{\prime\prime})}e^{-\frac{|t^\prime-t^{\prime\prime}|}{\tau_A}}$

\noindent Let's take $t_1>t_2$ for calculating $K$ 
\begin{equation}
\begin{split}
K &=\int_0^{t_1} dt^\prime\int_0^{t_2} dt^{\prime\prime} e^{\frac{k}{\gamma} (t^\prime+t^{\prime\prime})}e^{-\frac{|t^\prime-t^{\prime\prime}|}{\tau_A}}\\
&=\int_{t_2}^{t_1} dt^\prime\int_0^{t_2} dt^{\prime\prime}e^{\frac{k}{\gamma} (t^\prime+t^{\prime\prime})}e^{-\frac{|t^\prime-t^{\prime\prime}|}{\tau_A}}+\int_0^{t_2} dt^\prime\int_0^{t_2} dt^{\prime\prime}e^{\frac{k}{\gamma} (t^\prime+t^{\prime\prime})}e^{-\frac{|t^\prime-t^{\prime\prime}|}{\tau_A}}\\
&=\int_{t_2}^{t_1} dt^\prime\int_0^{t_2} dt^{\prime\prime}e^{\frac{k}{\gamma} (t^\prime+t^{\prime\prime})}e^{-\frac{(t^\prime-t^{\prime\prime})}{\tau_A}}+2\int_0^{t_2} dt^\prime\int_0^{t^\prime} dt^{\prime\prime}e^{\frac{k}{\gamma} (t^\prime+t^{\prime\prime})}e^{-\frac{(t^\prime-t^{\prime\prime})}{\tau_A}}\\
&=\frac{e^{(\frac{k}{\gamma}-\frac{1}{\tau_A})t_1+(\frac{k}{\gamma}+\frac{1}{\tau_A})t_2}-e^{2\frac{k}{\gamma}t_2}-e^{(\frac{k}{\gamma}-\frac{1}{\tau_A})t_1}-e^{(\frac{k}{\gamma}-\frac{1}{\tau_A})t_2}}{(\frac{k}{\gamma}-\frac{1}{\tau_A})(\frac{k}{\gamma}+\frac{1}{\tau_A})}+\frac{e^{2\frac{k}{\gamma}t_2}}{\frac{k}{\gamma}(\frac{k}{\gamma}+\frac{1}{\tau_A})}+\frac{1}{\frac{k}{\gamma}\left(\frac{k}{\gamma}-\frac{1}{\tau_A}\right)}
\label{eq:K}
\end{split}
\end{equation}
Substituting Eq.(\ref{eq:K}) in Eq.(\ref{eq:h(t1t2)}), we get
\begin{equation}
\begin{split}
h(t_1,t_2)&=\frac{k_BT}{k}e^{-\frac{k}{\gamma} (t_1-t_2)}+\frac{C}{\gamma^2}e^{-\frac{k}{\gamma} (t_1+t_2)} \times K\\
&=\frac{k_BT}{k}e^{-\frac{k}{\gamma} (t_1-t_2)}+\frac{C}{\gamma^2}e^{-\frac{k}{\gamma} (t_1+t_2)}\Bigg[\frac{e^{(\frac{k}{\gamma}-\frac{1}{\tau_A})t_1+(\frac{k}{\gamma}+\frac{1}{\tau_A})t_2}-e^{2\frac{k}{\gamma}t_2}-e^{(\frac{k}{\gamma}-\frac{1}{\tau_A})t_1}}{(\frac{k}{\gamma}-\frac{1}{\tau_A})(\frac{k}{\gamma}+\frac{1}{\tau_A})}\\
&-\frac{e^{(\frac{k}{\gamma}-\frac{1}{\tau_A})t_2}}{(\frac{k}{\gamma}-\frac{1}{\tau_A})(\frac{k}{\gamma}+\frac{1}{\tau_A})}+\frac{e^{2\frac{k}{\gamma}t_2}}{\frac{k}{\gamma}(\frac{k}{\gamma}+\frac{1}{\tau_A})}+\frac{1}{\frac{k}{\gamma}\left(\frac{k}{\gamma}-\frac{1}{\tau_A}\right)}\Bigg]\\
&=\frac{k_BT}{k}e^{-\frac{k}{\gamma} (t_1-t_2)}+\frac{C}{\gamma^2}\Bigg[\frac{e^{-\frac{1}{\tau_A}(t_1-t_2)}-e^{-\frac{k}{\gamma}(t_1-t_2)}-e^{-(\frac{k}{\gamma}t_2+\frac{1}{\tau_A}t_1)}-e^{-(\frac{k}{\gamma}t_1+\frac{1}{\tau_A}t_2)}}{(\frac{k}{\gamma}-\frac{1}{\tau_A})(\frac{k}{\gamma}+\frac{1}{\tau_A})}\\
&+\frac{e^{-\frac{k}{\gamma}(t_1-t_2)}}{\frac{k}{\gamma}(\frac{k}{\gamma}+\frac{1}{\tau_A})}+\frac{e^{-\frac{k}{\gamma} (t_1+t_2)}}{\frac{k}{\gamma}\left(\frac{k}{\gamma}-\frac{1}{\tau_A}\right)}\Bigg]
\end{split}
\end{equation}
\subsubsection{Calculation of variance of work $\left(\sigma_{W}^2\right)$ for transient work fluctuation theorem}
\begin{equation}
\begin{split}
\sigma_{W}^2 &=k^2u^2\int_{0}^{t} dt_1\int_{0}^{t} dt_2 h(t_1,t_2)\\
&=k^2u^2\int_{0}^{t} dt_1\int_{0}^{t} dt_2\Bigg[\frac{k_BT}{k}e^{-\frac{k}{\gamma} (t_1-t_2)}\\
&+\frac{C}{\gamma^2}\Bigg(\frac{e^{-\frac{1}{\tau_A}(t_1-t_2)}-e^{-\frac{k}{\gamma}(t_1-t_2)}-e^{-(\frac{k}{\gamma}t_2+\frac{1}{\tau_A}t_1)}-e^{-(\frac{k}{\gamma}t_1+\frac{1}{\tau_A}t_2)}}{(\frac{k}{\gamma}-\frac{1}{\tau_A})(\frac{k}{\gamma}+\frac{1}{\tau_A})}\\
&+\frac{e^{-\frac{k}{\gamma}(t_1-t_2)}}{\frac{k}{\gamma}(\frac{k}{\gamma}+\frac{1}{\tau_A})}+\frac{e^{-\frac{k}{\gamma} (t_1+t_2)}}{\frac{k}{\gamma}\left(\frac{k}{\gamma}-\frac{1}{\tau_A}\right)}\Bigg)\Bigg]\\
&=2k_BTku^2\int_{0}^{t} dt_1\int_{0}^{t_1} dt_2 e^{-\frac{k}{\gamma} (t_1-t_2)}\\
&+\frac{2k^2u^2C}{\gamma^2}\int_{0}^{t} dt_1\int_{0}^{t_1} dt_2\Bigg(\frac{e^{-\frac{1}{\tau_A}(t_1-t_2)}-e^{-\frac{k}{\gamma}(t_1-t_2)}-e^{-(\frac{k}{\gamma}t_2+\frac{1}{\tau_A}t_1)}-e^{-(\frac{k}{\gamma}t_1+\frac{1}{\tau_A}t_2)}}{(\frac{k}{\gamma}-\frac{1}{\tau_A})(\frac{k}{\gamma}+\frac{1}{\tau_A})}\\
&+\frac{e^{-\frac{k}{\gamma}(t_1-t_2)}}{\frac{k}{\gamma}(\frac{k}{\gamma}+\frac{1}{\tau_A})}+\frac{e^{-\frac{k}{\gamma} (t_1+t_2)}}{\frac{k}{\gamma}\left(\frac{k}{\gamma}-\frac{1}{\tau_A}\right)}\Bigg)\\
\label{eq:sigma}
\end{split}
\end{equation}
After integrating Eq.(\ref{eq:sigma}) and substituting the values of $\left<W\right>$ and $T_{eff}$, we get 
\begin{equation}
\begin{split}
\sigma_{W}^2 &=2k_BT_{eff}\left<W\right>+\frac{2k^2u^2C}{\gamma^2 (\frac{k}{\gamma}-\frac{1}{\tau_A})(\frac{k}{\gamma}+\frac{1}{\tau_A})}\Bigg[\frac{\gamma\tau_A}{k}\left(\frac{k}{\gamma}-\frac{1}{\tau_A}\right)t+\left(\tau_A^2e^{-\frac{1}{\tau_A}t}-\frac{\gamma^2}{k^2}e^{-\frac{k}{\gamma} t}\right)\\
&-\left(\tau_A^2-\frac{\gamma^2}{k^2}\right)-\frac{\gamma\tau_A}{k}e^{-(\frac{k}{\gamma}+\frac{1}{\tau_A})t}+\frac{\gamma\tau_A}{k}\left(e^{-\frac{1}{\tau_A}t}+e^{-\frac{k}{\gamma}t}\right)-\frac{\gamma\tau_A}{k}\Bigg]\\
&+\frac{\gamma}{k}\frac{u^2C} {(\frac{k}{\gamma}-\frac{1}{\tau_A})}\left(e^{-2\frac{k}{\gamma} t}-1\right)-\frac{\gamma}{k}\frac{2u^2C} {(\frac{k}{\gamma}-\frac{1}{\tau_A})}\left(e^{-\frac{k}{\gamma} t}-1\right)\\
\end{split}
\end{equation}
\subsubsection{Calculation of variance of work $\left(\sigma_{W_{ss}}^2\right)$for Steady state work fluctuation theorem}
\begin{equation}
\begin{split}
\textrm{Now,}\,\,\,\,\sigma_{W_{ss}}^2 &=k^2u^2\int_{0}^{t} dt_1\int_{0}^{t} dt_2 h(t_1+t_0,t_2+t_0)\\
&=k^2u^2\int_{0}^{t} dt_1\int_{0}^{t} dt_2\left[\frac{k_BT}{k}e^{-\frac{k}{\gamma} (t_1-t_2)}+\frac{k_BT_{act}}{ {(\frac{k}{\gamma}-\frac{1}{\tau_A})}} \left(\frac{e^{-\frac{1}{\tau_A}(t_1-t_2)}}{\gamma} -\frac{e^{-\frac{k}{\gamma}(t_1-t_2)}}{k\tau_A}\right)\right]\\
&=2k^2u^2\int_{0}^{t} dt_2\int_{0}^{t_2} dt_1 \left(\frac{k_BT}{k}e^{-\frac{k}{\gamma} (t_1-t_2)}\right)+\frac{2k^2u^2k_BT_{act}}{ {(\frac{k}{\gamma}-\frac{1}{\tau_A})}}\Bigg[ \frac{1}{\gamma}\int_{0}^{t} dt_1\int_{0}^{t_1} dt_2 e^{-\frac{1}{\tau_A}(t_1-t_2)}\\
&-\frac{1}{k\tau_A} \int_{0}^{t} dt_1\int_{0}^{t_1} dt_2 e^{-\frac{k}{\gamma}(t_1-t_2)}\Bigg]\\
&=2k_BT u^2\gamma t \left[1-\frac{\gamma}{kt}\left(e^{-\frac{k}{\gamma}t}-1\right)\right]+\frac{2k^2u^2 \tau_A k_BT_{act}t}{ {\gamma(\frac{k}{\gamma}-\frac{1}{\tau_A})}}\left[1-\frac{1}{t\tau_A} \left(e^{-\frac{1}{\tau_A}t}-1\right)\right] \\
&-\frac{2u^2 \gamma k_BT_{act}t}{ {\tau_A(\frac{k}{\gamma}-\frac{1}{\tau_A})}} \left[1-\frac{\gamma}{kt}\left(e^{-\frac{k}{\gamma}t}-1\right)\right] \Bigg] 
\end{split}
\end{equation}

\bibliographystyle{apsrev}
%\bibliography{active_fluctuation_physica_a_accepted}

\begin{thebibliography}{76}
\expandafter\ifx\csname natexlab\endcsname\relax\def\natexlab#1{#1}\fi
\expandafter\ifx\csname bibnamefont\endcsname\relax
  \def\bibnamefont#1{#1}\fi
\expandafter\ifx\csname bibfnamefont\endcsname\relax
  \def\bibfnamefont#1{#1}\fi
\expandafter\ifx\csname citenamefont\endcsname\relax
  \def\citenamefont#1{#1}\fi
\expandafter\ifx\csname url\endcsname\relax
  \def\url#1{\texttt{#1}}\fi
\expandafter\ifx\csname urlprefix\endcsname\relax\def\urlprefix{URL }\fi
\providecommand{\bibinfo}[2]{#2}
\providecommand{\eprint}[2][]{\url{#2}}

\bibitem[{\citenamefont{Zwanzig}(2001)}]{zwanzig2001nonequilibrium}
\bibinfo{author}{\bibfnamefont{R.}~\bibnamefont{Zwanzig}},
  \emph{\bibinfo{title}{Nonequilibrium Statistical Mechanics}}
  (\bibinfo{publisher}{Oxford University Press}, \bibinfo{year}{2001}).

\bibitem[{\citenamefont{Argun et~al.}(2016)\citenamefont{Argun, Moradi,
  Pin{\c{c}}e, Bagci, Imparato, and Volpe}}]{argun2016non}
\bibinfo{author}{\bibfnamefont{A.}~\bibnamefont{Argun}},
  \bibinfo{author}{\bibfnamefont{A.-R.} \bibnamefont{Moradi}},
  \bibinfo{author}{\bibfnamefont{E.}~\bibnamefont{Pin{\c{c}}e}},
  \bibinfo{author}{\bibfnamefont{G.~B.} \bibnamefont{Bagci}},
  \bibinfo{author}{\bibfnamefont{A.}~\bibnamefont{Imparato}}, \bibnamefont{and}
  \bibinfo{author}{\bibfnamefont{G.}~\bibnamefont{Volpe}},
  \bibinfo{journal}{Phys. Rev. E} \textbf{\bibinfo{volume}{94}},
  \bibinfo{pages}{062150} (\bibinfo{year}{2016}).

\bibitem[{\citenamefont{Krishnamurthy et~al.}(2016)\citenamefont{Krishnamurthy,
  Ghosh, Chatterji, Ganapathy, and Sood}}]{krishnamurthy2016micrometre}
\bibinfo{author}{\bibfnamefont{S.}~\bibnamefont{Krishnamurthy}},
  \bibinfo{author}{\bibfnamefont{S.}~\bibnamefont{Ghosh}},
  \bibinfo{author}{\bibfnamefont{D.}~\bibnamefont{Chatterji}},
  \bibinfo{author}{\bibfnamefont{R.}~\bibnamefont{Ganapathy}},
  \bibnamefont{and} \bibinfo{author}{\bibfnamefont{A.~K.} \bibnamefont{Sood}},
  \bibinfo{journal}{Nat. Phys.} \textbf{\bibinfo{volume}{12}},
  \bibinfo{pages}{1134} (\bibinfo{year}{2016}).

\bibitem[{\citenamefont{Maggi et~al.}(2014)\citenamefont{Maggi, Paoluzzi,
  Pellicciotta, Lepore, Angelani, and Di~Leonardo}}]{maggi2014generalized}
\bibinfo{author}{\bibfnamefont{C.}~\bibnamefont{Maggi}},
  \bibinfo{author}{\bibfnamefont{M.}~\bibnamefont{Paoluzzi}},
  \bibinfo{author}{\bibfnamefont{N.}~\bibnamefont{Pellicciotta}},
  \bibinfo{author}{\bibfnamefont{A.}~\bibnamefont{Lepore}},
  \bibinfo{author}{\bibfnamefont{L.}~\bibnamefont{Angelani}}, \bibnamefont{and}
  \bibinfo{author}{\bibfnamefont{R.}~\bibnamefont{Di~Leonardo}},
  \bibinfo{journal}{Phys. Rev. Lett.} \textbf{\bibinfo{volume}{113}},
  \bibinfo{pages}{238303} (\bibinfo{year}{2014}).

\bibitem[{\citenamefont{Harder et~al.}(2014)\citenamefont{Harder, Valeriani,
  and Cacciuto}}]{harder2014activity}
\bibinfo{author}{\bibfnamefont{J.}~\bibnamefont{Harder}},
  \bibinfo{author}{\bibfnamefont{C.}~\bibnamefont{Valeriani}},
  \bibnamefont{and} \bibinfo{author}{\bibfnamefont{A.}~\bibnamefont{Cacciuto}},
  \bibinfo{journal}{Phys. Rev. E} \textbf{\bibinfo{volume}{90}},
  \bibinfo{pages}{062312} (\bibinfo{year}{2014}).

\bibitem[{\citenamefont{Samanta and Chakrabarti}(2016)}]{samanta2016chain}
\bibinfo{author}{\bibfnamefont{N.}~\bibnamefont{Samanta}} \bibnamefont{and}
  \bibinfo{author}{\bibfnamefont{R.}~\bibnamefont{Chakrabarti}},
  \bibinfo{journal}{J. Phys. A} \textbf{\bibinfo{volume}{49}},
  \bibinfo{pages}{195601} (\bibinfo{year}{2016}).

\bibitem[{\citenamefont{Vandebroek and
  Vanderzande}(2015)}]{vandebroek2015dynamics}
\bibinfo{author}{\bibfnamefont{H.}~\bibnamefont{Vandebroek}} \bibnamefont{and}
  \bibinfo{author}{\bibfnamefont{C.}~\bibnamefont{Vanderzande}},
  \bibinfo{journal}{Phys. Rev. E} \textbf{\bibinfo{volume}{92}},
  \bibinfo{pages}{060601} (\bibinfo{year}{2015}).

\bibitem[{\citenamefont{Kaiser and L{\"o}wen}(2014)}]{kaiser2014unusual}
\bibinfo{author}{\bibfnamefont{A.}~\bibnamefont{Kaiser}} \bibnamefont{and}
  \bibinfo{author}{\bibfnamefont{H.}~\bibnamefont{L{\"o}wen}},
  \bibinfo{journal}{J. Chem. Phys.} \textbf{\bibinfo{volume}{141}},
  \bibinfo{pages}{044903} (\bibinfo{year}{2014}).

\bibitem[{\citenamefont{Eisenstecken et~al.}(2016)\citenamefont{Eisenstecken,
  Gompper, and Winkler}}]{eisenstecken2016conformational}
\bibinfo{author}{\bibfnamefont{T.}~\bibnamefont{Eisenstecken}},
  \bibinfo{author}{\bibfnamefont{G.}~\bibnamefont{Gompper}}, \bibnamefont{and}
  \bibinfo{author}{\bibfnamefont{R.~G.} \bibnamefont{Winkler}},
  \bibinfo{journal}{Polymers} \textbf{\bibinfo{volume}{8}},
  \bibinfo{pages}{304} (\bibinfo{year}{2016}).

\bibitem[{\citenamefont{Eisenstecken et~al.}(2017)\citenamefont{Eisenstecken,
  Gompper, and Winkler}}]{eisenstecken2017internal}
\bibinfo{author}{\bibfnamefont{T.}~\bibnamefont{Eisenstecken}},
  \bibinfo{author}{\bibfnamefont{G.}~\bibnamefont{Gompper}}, \bibnamefont{and}
  \bibinfo{author}{\bibfnamefont{R.~G.} \bibnamefont{Winkler}},
  \bibinfo{journal}{J. Chem. Phys.} \textbf{\bibinfo{volume}{146}},
  \bibinfo{pages}{154903} (\bibinfo{year}{2017}).

\bibitem[{\citenamefont{Osmanovi{\'c} and Rabin}(2017)}]{osmanovic2017dynamics}
\bibinfo{author}{\bibfnamefont{D.}~\bibnamefont{Osmanovi{\'c}}}
  \bibnamefont{and} \bibinfo{author}{\bibfnamefont{Y.}~\bibnamefont{Rabin}},
  \bibinfo{journal}{Soft Matter} \textbf{\bibinfo{volume}{13}},
  \bibinfo{pages}{963} (\bibinfo{year}{2017}).

\bibitem[{\citenamefont{Suzuki et~al.}(2015)\citenamefont{Suzuki, Weber, Frey,
  and Bausch}}]{suzuki2015polar}
\bibinfo{author}{\bibfnamefont{R.}~\bibnamefont{Suzuki}},
  \bibinfo{author}{\bibfnamefont{C.~A.} \bibnamefont{Weber}},
  \bibinfo{author}{\bibfnamefont{E.}~\bibnamefont{Frey}}, \bibnamefont{and}
  \bibinfo{author}{\bibfnamefont{A.~R.} \bibnamefont{Bausch}},
  \bibinfo{journal}{Nat. Phys.} \textbf{\bibinfo{volume}{11}},
  \bibinfo{pages}{839} (\bibinfo{year}{2015}).

\bibitem[{\citenamefont{Brangwynne et~al.}(2008)\citenamefont{Brangwynne,
  Koenderink, MacKintosh, and Weitz}}]{brangwynne2008nonequilibrium}
\bibinfo{author}{\bibfnamefont{C.~P.} \bibnamefont{Brangwynne}},
  \bibinfo{author}{\bibfnamefont{G.~H.} \bibnamefont{Koenderink}},
  \bibinfo{author}{\bibfnamefont{F.~C.} \bibnamefont{MacKintosh}},
  \bibnamefont{and} \bibinfo{author}{\bibfnamefont{D.~A.} \bibnamefont{Weitz}},
  \bibinfo{journal}{Phys. Rev. Lett.} \textbf{\bibinfo{volume}{100}},
  \bibinfo{pages}{118104} (\bibinfo{year}{2008}).

\bibitem[{\citenamefont{Faris et~al.}(2009)\citenamefont{Faris, Lacoste,
  P{\'e}cr{\'e}aux, Joanny, Prost, and Bassereau}}]{faris2009membrane}
\bibinfo{author}{\bibfnamefont{M.~E.~A.} \bibnamefont{Faris}},
  \bibinfo{author}{\bibfnamefont{D.}~\bibnamefont{Lacoste}},
  \bibinfo{author}{\bibfnamefont{J.}~\bibnamefont{P{\'e}cr{\'e}aux}},
  \bibinfo{author}{\bibfnamefont{J.-F.} \bibnamefont{Joanny}},
  \bibinfo{author}{\bibfnamefont{J.}~\bibnamefont{Prost}}, \bibnamefont{and}
  \bibinfo{author}{\bibfnamefont{P.}~\bibnamefont{Bassereau}},
  \bibinfo{journal}{Phys. Rev. Lett.} \textbf{\bibinfo{volume}{102}},
  \bibinfo{pages}{038102} (\bibinfo{year}{2009}).

\bibitem[{\citenamefont{Shin et~al.}(2015)\citenamefont{Shin, Cherstvy, Kim,
  and Metzler}}]{shin2015facilitation}
\bibinfo{author}{\bibfnamefont{J.}~\bibnamefont{Shin}},
  \bibinfo{author}{\bibfnamefont{A.~G.} \bibnamefont{Cherstvy}},
  \bibinfo{author}{\bibfnamefont{W.~K.} \bibnamefont{Kim}}, \bibnamefont{and}
  \bibinfo{author}{\bibfnamefont{R.}~\bibnamefont{Metzler}},
  \bibinfo{journal}{New J. Phys.} \textbf{\bibinfo{volume}{17}},
  \bibinfo{pages}{113008} (\bibinfo{year}{2015}).

\bibitem[{\citenamefont{Shin et~al.}(2017)\citenamefont{Shin, Cherstvy, Kim,
  and Zaburdaev}}]{shin2017elasticity}
\bibinfo{author}{\bibfnamefont{J.}~\bibnamefont{Shin}},
  \bibinfo{author}{\bibfnamefont{A.~G.} \bibnamefont{Cherstvy}},
  \bibinfo{author}{\bibfnamefont{W.~K.} \bibnamefont{Kim}}, \bibnamefont{and}
  \bibinfo{author}{\bibfnamefont{V.}~\bibnamefont{Zaburdaev}},
  \bibinfo{journal}{Phys. Chem. Chem. Phys.} \textbf{\bibinfo{volume}{19}},
  \bibinfo{pages}{18338} (\bibinfo{year}{2017}).

\bibitem[{\citenamefont{Puglisi and Marconi}(2017)}]{marconi2017}
\bibinfo{author}{\bibfnamefont{A.}~\bibnamefont{Puglisi}} \bibnamefont{and}
  \bibinfo{author}{\bibfnamefont{U.~M.~B.} \bibnamefont{Marconi}},
  \bibinfo{journal}{Entropy} \textbf{\bibinfo{volume}{19}},
  \bibinfo{pages}{356} (\bibinfo{year}{2017}).

\bibitem[{\citenamefont{Marconi et~al.}(2017)\citenamefont{Marconi, Puglisi,
  and Maggi}}]{marconi2017heat}
\bibinfo{author}{\bibfnamefont{U.~M.~B.} \bibnamefont{Marconi}},
  \bibinfo{author}{\bibfnamefont{A.}~\bibnamefont{Puglisi}}, \bibnamefont{and}
  \bibinfo{author}{\bibfnamefont{C.}~\bibnamefont{Maggi}},
  \bibinfo{journal}{Scientific Reports} \textbf{\bibinfo{volume}{7}},
  \bibinfo{pages}{46496} (\bibinfo{year}{2017}).

\bibitem[{\citenamefont{Sevick et~al.}(2008)\citenamefont{Sevick, Prabhakar,
  Williams, and Searles}}]{sevick2008fluctuation}
\bibinfo{author}{\bibfnamefont{E.~M.} \bibnamefont{Sevick}},
  \bibinfo{author}{\bibfnamefont{R.}~\bibnamefont{Prabhakar}},
  \bibinfo{author}{\bibfnamefont{S.~R.} \bibnamefont{Williams}},
  \bibnamefont{and} \bibinfo{author}{\bibfnamefont{D.~J.}
  \bibnamefont{Searles}}, \bibinfo{journal}{Annu. Rev. Phys. Chem.}
  \textbf{\bibinfo{volume}{59}}, \bibinfo{pages}{603} (\bibinfo{year}{2008}).

\bibitem[{\citenamefont{Ritort}(2008)}]{ritort2008nonequilibrium}
\bibinfo{author}{\bibfnamefont{F.}~\bibnamefont{Ritort}},
  \bibinfo{journal}{Adv. Chem. Phys.} \textbf{\bibinfo{volume}{137}},
  \bibinfo{pages}{31} (\bibinfo{year}{2008}).

\bibitem[{\citenamefont{Mart{\'\i}nez et~al.}(2017)\citenamefont{Mart{\'\i}nez,
  Rold{\'a}n, Dinis, and Rica}}]{martinez2017colloidal}
\bibinfo{author}{\bibfnamefont{I.~A.} \bibnamefont{Mart{\'\i}nez}},
  \bibinfo{author}{\bibfnamefont{{\'E}.}~\bibnamefont{Rold{\'a}n}},
  \bibinfo{author}{\bibfnamefont{L.}~\bibnamefont{Dinis}}, \bibnamefont{and}
  \bibinfo{author}{\bibfnamefont{R.~A.} \bibnamefont{Rica}},
  \bibinfo{journal}{Soft Matter} \textbf{\bibinfo{volume}{13}},
  \bibinfo{pages}{22} (\bibinfo{year}{2017}).

\bibitem[{\citenamefont{Varghese et~al.}(2013)\citenamefont{Varghese,
  Vemparala, and Rajesh}}]{varghese2013force}
\bibinfo{author}{\bibfnamefont{A.}~\bibnamefont{Varghese}},
  \bibinfo{author}{\bibfnamefont{S.}~\bibnamefont{Vemparala}},
  \bibnamefont{and} \bibinfo{author}{\bibfnamefont{R.}~\bibnamefont{Rajesh}},
  \bibinfo{journal}{Phys. Rev. E} \textbf{\bibinfo{volume}{88}},
  \bibinfo{pages}{022134} (\bibinfo{year}{2013}).

\bibitem[{\citenamefont{Colangeli et~al.}(2011)\citenamefont{Colangeli, Klages,
  De~Gregorio, and Rondoni}}]{colangeli2011steady}
\bibinfo{author}{\bibfnamefont{M.}~\bibnamefont{Colangeli}},
  \bibinfo{author}{\bibfnamefont{R.}~\bibnamefont{Klages}},
  \bibinfo{author}{\bibfnamefont{P.}~\bibnamefont{De~Gregorio}},
  \bibnamefont{and} \bibinfo{author}{\bibfnamefont{L.}~\bibnamefont{Rondoni}},
  \bibinfo{journal}{J. Stat. Mech.} \textbf{\bibinfo{volume}{P04021}}
  (\bibinfo{year}{2011}).

\bibitem[{\citenamefont{Saha et~al.}(2011)\citenamefont{Saha, Bhattacharjee,
  and Chakraborty}}]{saha2011work}
\bibinfo{author}{\bibfnamefont{A.}~\bibnamefont{Saha}},
  \bibinfo{author}{\bibfnamefont{J.~K.} \bibnamefont{Bhattacharjee}},
  \bibnamefont{and}
  \bibinfo{author}{\bibfnamefont{S.}~\bibnamefont{Chakraborty}},
  \bibinfo{journal}{Phys. Rev. E} \textbf{\bibinfo{volume}{83}},
  \bibinfo{pages}{011104} (\bibinfo{year}{2011}).

\bibitem[{\citenamefont{Saha and Bhattacharjee}(2007)}]{saha2007asymmetry}
\bibinfo{author}{\bibfnamefont{A.}~\bibnamefont{Saha}} \bibnamefont{and}
  \bibinfo{author}{\bibfnamefont{J.}~\bibnamefont{Bhattacharjee}},
  \bibinfo{journal}{J. Phys. A} \textbf{\bibinfo{volume}{40}},
  \bibinfo{pages}{13269} (\bibinfo{year}{2007}).

\bibitem[{\citenamefont{Evans and Searles}(1994)}]{evans1994equilibrium}
\bibinfo{author}{\bibfnamefont{D.~J.} \bibnamefont{Evans}} \bibnamefont{and}
  \bibinfo{author}{\bibfnamefont{D.~J.} \bibnamefont{Searles}},
  \bibinfo{journal}{Phys. Rev. E} \textbf{\bibinfo{volume}{50}},
  \bibinfo{pages}{1645} (\bibinfo{year}{1994}).

\bibitem[{\citenamefont{Kurchan}(1998)}]{kurchan1998fluctuation}
\bibinfo{author}{\bibfnamefont{J.}~\bibnamefont{Kurchan}}, \bibinfo{journal}{J.
  Phys. A} \textbf{\bibinfo{volume}{31}}, \bibinfo{pages}{3719}
  (\bibinfo{year}{1998}).

\bibitem[{\citenamefont{Crooks}(1999)}]{crooks1999entropy}
\bibinfo{author}{\bibfnamefont{G.~E.} \bibnamefont{Crooks}},
  \bibinfo{journal}{Phys. Rev. E} \textbf{\bibinfo{volume}{60}},
  \bibinfo{pages}{2721} (\bibinfo{year}{1999}).

\bibitem[{\citenamefont{Gallavotti and Cohen}(1995)}]{gallavotti1995dynamical}
\bibinfo{author}{\bibfnamefont{G.}~\bibnamefont{Gallavotti}} \bibnamefont{and}
  \bibinfo{author}{\bibfnamefont{E.}~\bibnamefont{Cohen}},
  \bibinfo{journal}{Phys. Rev. Lett.} \textbf{\bibinfo{volume}{74}},
  \bibinfo{pages}{2694} (\bibinfo{year}{1995}).

\bibitem[{\citenamefont{Speck and Seifert}(2005)}]{speck2005integral}
\bibinfo{author}{\bibfnamefont{T.}~\bibnamefont{Speck}} \bibnamefont{and}
  \bibinfo{author}{\bibfnamefont{U.}~\bibnamefont{Seifert}},
  \bibinfo{journal}{J. Phys. A} \textbf{\bibinfo{volume}{38}},
  \bibinfo{pages}{L581} (\bibinfo{year}{2005}).

\bibitem[{\citenamefont{Van~Zon and Cohen}(2003)}]{van2003extension}
\bibinfo{author}{\bibfnamefont{R.}~\bibnamefont{Van~Zon}} \bibnamefont{and}
  \bibinfo{author}{\bibfnamefont{E.}~\bibnamefont{Cohen}},
  \bibinfo{journal}{Phys. Rev. Lett.} \textbf{\bibinfo{volume}{91}},
  \bibinfo{pages}{110601} (\bibinfo{year}{2003}).

\bibitem[{\citenamefont{Seifert}(2005)}]{seifert2005entropy}
\bibinfo{author}{\bibfnamefont{U.}~\bibnamefont{Seifert}},
  \bibinfo{journal}{Phys. Rev. Lett.} \textbf{\bibinfo{volume}{95}},
  \bibinfo{pages}{040602} (\bibinfo{year}{2005}).

\bibitem[{\citenamefont{Speck et~al.}(2007)\citenamefont{Speck, Blickle,
  Bechinger, and Seifert}}]{speck2007distribution}
\bibinfo{author}{\bibfnamefont{T.}~\bibnamefont{Speck}},
  \bibinfo{author}{\bibfnamefont{V.}~\bibnamefont{Blickle}},
  \bibinfo{author}{\bibfnamefont{C.}~\bibnamefont{Bechinger}},
  \bibnamefont{and} \bibinfo{author}{\bibfnamefont{U.}~\bibnamefont{Seifert}},
  \bibinfo{journal}{Euro. Phys. Lett.} \textbf{\bibinfo{volume}{79}},
  \bibinfo{pages}{30002} (\bibinfo{year}{2007}).

\bibitem[{\citenamefont{Garnier and
  Ciliberto}(2005)}]{garnier2005nonequilibrium}
\bibinfo{author}{\bibfnamefont{N.}~\bibnamefont{Garnier}} \bibnamefont{and}
  \bibinfo{author}{\bibfnamefont{S.}~\bibnamefont{Ciliberto}},
  \bibinfo{journal}{Phys. Rev. E} \textbf{\bibinfo{volume}{71}},
  \bibinfo{pages}{060101} (\bibinfo{year}{2005}).

\bibitem[{\citenamefont{Hummer and Szabo}(2001)}]{hummer2001free}
\bibinfo{author}{\bibfnamefont{G.}~\bibnamefont{Hummer}} \bibnamefont{and}
  \bibinfo{author}{\bibfnamefont{A.}~\bibnamefont{Szabo}},
  \bibinfo{journal}{Proc. Natl. Acad. Sci. USA} \textbf{\bibinfo{volume}{98}},
  \bibinfo{pages}{3658} (\bibinfo{year}{2001}).

\bibitem[{\citenamefont{Liphardt et~al.}(2002)\citenamefont{Liphardt, Dumont,
  Smith, Tinoco, and Bustamante}}]{liphardt2002equilibrium}
\bibinfo{author}{\bibfnamefont{J.}~\bibnamefont{Liphardt}},
  \bibinfo{author}{\bibfnamefont{S.}~\bibnamefont{Dumont}},
  \bibinfo{author}{\bibfnamefont{S.~B.} \bibnamefont{Smith}},
  \bibinfo{author}{\bibfnamefont{I.}~\bibnamefont{Tinoco}}, \bibnamefont{and}
  \bibinfo{author}{\bibfnamefont{C.}~\bibnamefont{Bustamante}},
  \bibinfo{journal}{Science} \textbf{\bibinfo{volume}{296}},
  \bibinfo{pages}{1832} (\bibinfo{year}{2002}).

\bibitem[{\citenamefont{Beck and Cohen}(2004)}]{beck2004superstatistical}
\bibinfo{author}{\bibfnamefont{C.}~\bibnamefont{Beck}} \bibnamefont{and}
  \bibinfo{author}{\bibfnamefont{E.}~\bibnamefont{Cohen}},
  \bibinfo{journal}{Physica A} \textbf{\bibinfo{volume}{344}},
  \bibinfo{pages}{393} (\bibinfo{year}{2004}).

\bibitem[{\citenamefont{Touchette and Cohen}(2007)}]{touchette2007fluctuation}
\bibinfo{author}{\bibfnamefont{H.}~\bibnamefont{Touchette}} \bibnamefont{and}
  \bibinfo{author}{\bibfnamefont{E.}~\bibnamefont{Cohen}},
  \bibinfo{journal}{Phys. Rev. E} \textbf{\bibinfo{volume}{76}},
  \bibinfo{pages}{020101} (\bibinfo{year}{2007}).

\bibitem[{\citenamefont{Sellitto}(2009)}]{sellitto2009fluctuation}
\bibinfo{author}{\bibfnamefont{M.}~\bibnamefont{Sellitto}},
  \bibinfo{journal}{Phys. Rev. E} \textbf{\bibinfo{volume}{80}},
  \bibinfo{pages}{011134} (\bibinfo{year}{2009}).

\bibitem[{\citenamefont{Chechkin and Klages}(2009)}]{chechkin2009fluctuation}
\bibinfo{author}{\bibfnamefont{A.~V.} \bibnamefont{Chechkin}} \bibnamefont{and}
  \bibinfo{author}{\bibfnamefont{R.}~\bibnamefont{Klages}},
  \bibinfo{journal}{J. Stat. Mech.} \textbf{\bibinfo{volume}{L03002}}
  (\bibinfo{year}{2009}).

\bibitem[{\citenamefont{Harris and Touchette}(2009)}]{harris2009current}
\bibinfo{author}{\bibfnamefont{R.}~\bibnamefont{Harris}} \bibnamefont{and}
  \bibinfo{author}{\bibfnamefont{H.}~\bibnamefont{Touchette}},
  \bibinfo{journal}{J. Phys. A} \textbf{\bibinfo{volume}{42}},
  \bibinfo{pages}{342001} (\bibinfo{year}{2009}).

\bibitem[{\citenamefont{Budini}(2012)}]{budini2012generalized}
\bibinfo{author}{\bibfnamefont{A.~A.} \bibnamefont{Budini}},
  \bibinfo{journal}{Phys. Rev. E} \textbf{\bibinfo{volume}{86}},
  \bibinfo{pages}{011109} (\bibinfo{year}{2012}).

\bibitem[{\citenamefont{Zamponi
  et~al.}(2005{\natexlab{a}})\citenamefont{Zamponi, Ruocco, and
  Angelani}}]{zamponi2005generalized}
\bibinfo{author}{\bibfnamefont{F.}~\bibnamefont{Zamponi}},
  \bibinfo{author}{\bibfnamefont{G.}~\bibnamefont{Ruocco}}, \bibnamefont{and}
  \bibinfo{author}{\bibfnamefont{L.}~\bibnamefont{Angelani}},
  \bibinfo{journal}{Phys. Rev. E} \textbf{\bibinfo{volume}{71}},
  \bibinfo{pages}{020101} (\bibinfo{year}{2005}{\natexlab{a}}).

\bibitem[{\citenamefont{Zamponi
  et~al.}(2005{\natexlab{b}})\citenamefont{Zamponi, Bonetto, Cugliandolo, and
  Kurchan}}]{zamponi2005fluctuation}
\bibinfo{author}{\bibfnamefont{F.}~\bibnamefont{Zamponi}},
  \bibinfo{author}{\bibfnamefont{F.}~\bibnamefont{Bonetto}},
  \bibinfo{author}{\bibfnamefont{L.~F.} \bibnamefont{Cugliandolo}},
  \bibnamefont{and} \bibinfo{author}{\bibfnamefont{J.}~\bibnamefont{Kurchan}},
  \bibinfo{journal}{J. Stat. Mech.} \textbf{\bibinfo{volume}{P09013}}
  (\bibinfo{year}{2005}{\natexlab{b}}).

\bibitem[{\citenamefont{Chechkin et~al.}(2012)\citenamefont{Chechkin, Lenz, and
  Klages}}]{chechkin2012normal}
\bibinfo{author}{\bibfnamefont{A.~V.} \bibnamefont{Chechkin}},
  \bibinfo{author}{\bibfnamefont{F.}~\bibnamefont{Lenz}}, \bibnamefont{and}
  \bibinfo{author}{\bibfnamefont{R.}~\bibnamefont{Klages}},
  \bibinfo{journal}{J. Stat. Mech.} \textbf{\bibinfo{volume}{L11001}}
  (\bibinfo{year}{2012}).

\bibitem[{\citenamefont{Seifert}(2011)}]{seifert2011stochastic}
\bibinfo{author}{\bibfnamefont{U.}~\bibnamefont{Seifert}},
  \bibinfo{journal}{Euro. Phys. J. E} \textbf{\bibinfo{volume}{34}},
  \bibinfo{pages}{1} (\bibinfo{year}{2011}).

\bibitem[{\citenamefont{Lacoste and Mallick}(2009)}]{lacoste2009fluctuation}
\bibinfo{author}{\bibfnamefont{D.}~\bibnamefont{Lacoste}} \bibnamefont{and}
  \bibinfo{author}{\bibfnamefont{K.}~\bibnamefont{Mallick}},
  \bibinfo{journal}{Phys. Rev. E} \textbf{\bibinfo{volume}{80}},
  \bibinfo{pages}{021923} (\bibinfo{year}{2009}).

\bibitem[{\citenamefont{Speck}(2016)}]{speck2016stochastic}
\bibinfo{author}{\bibfnamefont{T.}~\bibnamefont{Speck}},
  \bibinfo{journal}{Europhys. Lett} \textbf{\bibinfo{volume}{114}},
  \bibinfo{pages}{30006} (\bibinfo{year}{2016}).

\bibitem[{\citenamefont{Chaudhuri}(2014)}]{chaudhuri2014active}
\bibinfo{author}{\bibfnamefont{D.}~\bibnamefont{Chaudhuri}},
  \bibinfo{journal}{Phys. Rev. E} \textbf{\bibinfo{volume}{90}},
  \bibinfo{pages}{022131} (\bibinfo{year}{2014}).

\bibitem[{\citenamefont{Pietzonka and Seifert}(2017)}]{pietzonka2017entropy}
\bibinfo{author}{\bibfnamefont{P.}~\bibnamefont{Pietzonka}} \bibnamefont{and}
  \bibinfo{author}{\bibfnamefont{U.}~\bibnamefont{Seifert}},
  \bibinfo{journal}{J. Phys. A} \textbf{\bibinfo{volume}{51}},
  \bibinfo{pages}{01LT01} (\bibinfo{year}{2017}).

\bibitem[{\citenamefont{Mandal et~al.}(2017)\citenamefont{Mandal, Klymko, and
  DeWeese}}]{mandal2017entropy}
\bibinfo{author}{\bibfnamefont{D.}~\bibnamefont{Mandal}},
  \bibinfo{author}{\bibfnamefont{K.}~\bibnamefont{Klymko}}, \bibnamefont{and}
  \bibinfo{author}{\bibfnamefont{M.~R.} \bibnamefont{DeWeese}},
  \bibinfo{journal}{Phys. Rev. Lett.} \textbf{\bibinfo{volume}{119}},
  \bibinfo{pages}{258001} (\bibinfo{year}{2017}).

\bibitem[{\citenamefont{Jarzynski}(1997{\natexlab{a}})}]{jarzynski1997nonequilibrium}
\bibinfo{author}{\bibfnamefont{C.}~\bibnamefont{Jarzynski}},
  \bibinfo{journal}{Phys. Rev. Lett.} \textbf{\bibinfo{volume}{78}},
  \bibinfo{pages}{2690} (\bibinfo{year}{1997}{\natexlab{a}}).

\bibitem[{\citenamefont{Jarzynski}(1997{\natexlab{b}})}]{jarzynski1997equilibrium}
\bibinfo{author}{\bibfnamefont{C.}~\bibnamefont{Jarzynski}},
  \bibinfo{journal}{Phys. Rev. E} \textbf{\bibinfo{volume}{56}},
  \bibinfo{pages}{5018} (\bibinfo{year}{1997}{\natexlab{b}}).

\bibitem[{\citenamefont{Wu and Libchaber}(2000)}]{wu2000particle}
\bibinfo{author}{\bibfnamefont{X.-L.} \bibnamefont{Wu}} \bibnamefont{and}
  \bibinfo{author}{\bibfnamefont{A.}~\bibnamefont{Libchaber}},
  \bibinfo{journal}{Phys. Rev. Lett.} \textbf{\bibinfo{volume}{84}},
  \bibinfo{pages}{3017} (\bibinfo{year}{2000}).

\bibitem[{\citenamefont{Vandebroek and
  Vanderzande}(2017)}]{vandebroek2017effect}
\bibinfo{author}{\bibfnamefont{H.}~\bibnamefont{Vandebroek}} \bibnamefont{and}
  \bibinfo{author}{\bibfnamefont{C.}~\bibnamefont{Vanderzande}},
  \bibinfo{journal}{Soft Matter} \textbf{\bibinfo{volume}{13}},
  \bibinfo{pages}{2181} (\bibinfo{year}{2017}).

\bibitem[{\citenamefont{Berthier and Kurchan}(2013)}]{berthier2013non}
\bibinfo{author}{\bibfnamefont{L.}~\bibnamefont{Berthier}} \bibnamefont{and}
  \bibinfo{author}{\bibfnamefont{J.}~\bibnamefont{Kurchan}},
  \bibinfo{journal}{Nat. Phys.} \textbf{\bibinfo{volume}{9}},
  \bibinfo{pages}{310} (\bibinfo{year}{2013}).

\bibitem[{\citenamefont{Stuhrmann et~al.}(2012)\citenamefont{Stuhrmann,
  e~Silva, Depken, MacKintosh, and Koenderink}}]{stuhrmann2012nonequilibrium}
\bibinfo{author}{\bibfnamefont{B.}~\bibnamefont{Stuhrmann}},
  \bibinfo{author}{\bibfnamefont{M.~S.} \bibnamefont{e~Silva}},
  \bibinfo{author}{\bibfnamefont{M.}~\bibnamefont{Depken}},
  \bibinfo{author}{\bibfnamefont{F.~C.} \bibnamefont{MacKintosh}},
  \bibnamefont{and} \bibinfo{author}{\bibfnamefont{G.~H.}
  \bibnamefont{Koenderink}}, \bibinfo{journal}{Phys. Rev. E}
  \textbf{\bibinfo{volume}{86}}, \bibinfo{pages}{020901}
  (\bibinfo{year}{2012}).

\bibitem[{\citenamefont{Toyota et~al.}(2011)\citenamefont{Toyota, Head,
  Schmidt, and Mizuno}}]{toyota2011non}
\bibinfo{author}{\bibfnamefont{T.}~\bibnamefont{Toyota}},
  \bibinfo{author}{\bibfnamefont{D.~A.} \bibnamefont{Head}},
  \bibinfo{author}{\bibfnamefont{C.~F.} \bibnamefont{Schmidt}},
  \bibnamefont{and} \bibinfo{author}{\bibfnamefont{D.}~\bibnamefont{Mizuno}},
  \bibinfo{journal}{Soft Matter} \textbf{\bibinfo{volume}{7}},
  \bibinfo{pages}{3234} (\bibinfo{year}{2011}).

\bibitem[{\citenamefont{Sonn-Segev et~al.}(2017)\citenamefont{Sonn-Segev,
  Bernheim-Groswasser, and Roichman}}]{sonn2017scale}
\bibinfo{author}{\bibfnamefont{A.}~\bibnamefont{Sonn-Segev}},
  \bibinfo{author}{\bibfnamefont{A.}~\bibnamefont{Bernheim-Groswasser}},
  \bibnamefont{and} \bibinfo{author}{\bibfnamefont{Y.}~\bibnamefont{Roichman}},
  \bibinfo{journal}{Soft Matter} \textbf{\bibinfo{volume}{13}},
  \bibinfo{pages}{7352} (\bibinfo{year}{2017}).

\bibitem[{\citenamefont{Ghosh and Gov}(2014)}]{ghosh2014dynamics}
\bibinfo{author}{\bibfnamefont{A.}~\bibnamefont{Ghosh}} \bibnamefont{and}
  \bibinfo{author}{\bibfnamefont{N.}~\bibnamefont{Gov}},
  \bibinfo{journal}{Biophys. J.} \textbf{\bibinfo{volume}{107}},
  \bibinfo{pages}{1065} (\bibinfo{year}{2014}).

\bibitem[{\citenamefont{Zakine et~al.}(2017)\citenamefont{Zakine, Solon,
  Gingrich, and van Wijland}}]{zakine2017stochastic}
\bibinfo{author}{\bibfnamefont{R.}~\bibnamefont{Zakine}},
  \bibinfo{author}{\bibfnamefont{A.}~\bibnamefont{Solon}},
  \bibinfo{author}{\bibfnamefont{T.}~\bibnamefont{Gingrich}}, \bibnamefont{and}
  \bibinfo{author}{\bibfnamefont{F.}~\bibnamefont{van Wijland}},
  \bibinfo{journal}{Entropy} \textbf{\bibinfo{volume}{19}},
  \bibinfo{pages}{193} (\bibinfo{year}{2017}).

\bibitem[{\citenamefont{Kanazawa et~al.}(2012)\citenamefont{Kanazawa, Sagawa,
  and Hayakawa}}]{kanazawa2012stochastic}
\bibinfo{author}{\bibfnamefont{K.}~\bibnamefont{Kanazawa}},
  \bibinfo{author}{\bibfnamefont{T.}~\bibnamefont{Sagawa}}, \bibnamefont{and}
  \bibinfo{author}{\bibfnamefont{H.}~\bibnamefont{Hayakawa}},
  \bibinfo{journal}{Phys. Rev. Lett.} \textbf{\bibinfo{volume}{108}},
  \bibinfo{pages}{210601} (\bibinfo{year}{2012}).

\bibitem[{\citenamefont{Wulfert et~al.}(2017)\citenamefont{Wulfert, Oechsle,
  Speck, and Seifert}}]{wulfert2017driven}
\bibinfo{author}{\bibfnamefont{R.}~\bibnamefont{Wulfert}},
  \bibinfo{author}{\bibfnamefont{M.}~\bibnamefont{Oechsle}},
  \bibinfo{author}{\bibfnamefont{T.}~\bibnamefont{Speck}}, \bibnamefont{and}
  \bibinfo{author}{\bibfnamefont{U.}~\bibnamefont{Seifert}},
  \bibinfo{journal}{Phys. Rev. E} \textbf{\bibinfo{volume}{95}},
  \bibinfo{pages}{050103} (\bibinfo{year}{2017}).

\bibitem[{\citenamefont{Dieterich et~al.}(2015)\citenamefont{Dieterich,
  Camunas-Soler, Ribezzi-Crivellari, Seifert, and
  Ritort}}]{dieterich2015single}
\bibinfo{author}{\bibfnamefont{E.}~\bibnamefont{Dieterich}},
  \bibinfo{author}{\bibfnamefont{J.}~\bibnamefont{Camunas-Soler}},
  \bibinfo{author}{\bibfnamefont{M.}~\bibnamefont{Ribezzi-Crivellari}},
  \bibinfo{author}{\bibfnamefont{U.}~\bibnamefont{Seifert}}, \bibnamefont{and}
  \bibinfo{author}{\bibfnamefont{F.}~\bibnamefont{Ritort}},
  \bibinfo{journal}{Nat. Phys.} \textbf{\bibinfo{volume}{11}},
  \bibinfo{pages}{971} (\bibinfo{year}{2015}).

\bibitem[{\citenamefont{Szamel}(2014)}]{szamel2014self}
\bibinfo{author}{\bibfnamefont{G.}~\bibnamefont{Szamel}},
  \bibinfo{journal}{Phys. Rev. E} \textbf{\bibinfo{volume}{90}},
  \bibinfo{pages}{012111} (\bibinfo{year}{2014}).

\bibitem[{\citenamefont{Sekimoto}(1998)}]{sekimoto1998langevin}
\bibinfo{author}{\bibfnamefont{K.}~\bibnamefont{Sekimoto}},
  \bibinfo{journal}{Prog. Theo. Phys. Suppl.} \textbf{\bibinfo{volume}{130}},
  \bibinfo{pages}{17} (\bibinfo{year}{1998}).

\bibitem[{\citenamefont{Saha et~al.}(2009)\citenamefont{Saha, Lahiri, and
  Jayannavar}}]{saha2009entropy}
\bibinfo{author}{\bibfnamefont{A.}~\bibnamefont{Saha}},
  \bibinfo{author}{\bibfnamefont{S.}~\bibnamefont{Lahiri}}, \bibnamefont{and}
  \bibinfo{author}{\bibfnamefont{A.}~\bibnamefont{Jayannavar}},
  \bibinfo{journal}{Phys. Rev. E} \textbf{\bibinfo{volume}{80}},
  \bibinfo{pages}{011117} (\bibinfo{year}{2009}).

\bibitem[{\citenamefont{Ghosh and Chaudhury}(2017)}]{chaudhury2017}
\bibinfo{author}{\bibfnamefont{B.}~\bibnamefont{Ghosh}} \bibnamefont{and}
  \bibinfo{author}{\bibfnamefont{S.}~\bibnamefont{Chaudhury}},
  \bibinfo{journal}{Physica A} \textbf{\bibinfo{volume}{466}},
  \bibinfo{pages}{133} (\bibinfo{year}{2017}).

\bibitem[{\citenamefont{Van~Zon and Cohen}(2004)}]{van2004extended}
\bibinfo{author}{\bibfnamefont{R.}~\bibnamefont{Van~Zon}} \bibnamefont{and}
  \bibinfo{author}{\bibfnamefont{E.}~\bibnamefont{Cohen}},
  \bibinfo{journal}{Phys. Rev. E} \textbf{\bibinfo{volume}{69}},
  \bibinfo{pages}{056121} (\bibinfo{year}{2004}).

\bibitem[{\citenamefont{Dhar}(2005)}]{dhar2005work}
\bibinfo{author}{\bibfnamefont{A.}~\bibnamefont{Dhar}}, \bibinfo{journal}{Phys.
  Rev. E} \textbf{\bibinfo{volume}{71}}, \bibinfo{pages}{036126}
  (\bibinfo{year}{2005}).

\bibitem[{\citenamefont{Trepagnier et~al.}(2004)\citenamefont{Trepagnier,
  Jarzynski, Ritort, Crooks, Bustamante, and
  Liphardt}}]{trepagnier2004experimental}
\bibinfo{author}{\bibfnamefont{E.}~\bibnamefont{Trepagnier}},
  \bibinfo{author}{\bibfnamefont{C.}~\bibnamefont{Jarzynski}},
  \bibinfo{author}{\bibfnamefont{F.}~\bibnamefont{Ritort}},
  \bibinfo{author}{\bibfnamefont{G.~E.} \bibnamefont{Crooks}},
  \bibinfo{author}{\bibfnamefont{C.}~\bibnamefont{Bustamante}},
  \bibnamefont{and} \bibinfo{author}{\bibfnamefont{J.}~\bibnamefont{Liphardt}},
  \bibinfo{journal}{Proc. Natl. Acad. Sci. USA} \textbf{\bibinfo{volume}{101}},
  \bibinfo{pages}{15038} (\bibinfo{year}{2004}).

\bibitem[{\citenamefont{Narayan and Dhar}(2003)}]{narayan2003reexamination}
\bibinfo{author}{\bibfnamefont{O.}~\bibnamefont{Narayan}} \bibnamefont{and}
  \bibinfo{author}{\bibfnamefont{A.}~\bibnamefont{Dhar}}, \bibinfo{journal}{J.
  Phys. A} \textbf{\bibinfo{volume}{37}}, \bibinfo{pages}{63}
  (\bibinfo{year}{2003}).

\bibitem[{\citenamefont{Chakrabarti}(2009)}]{chakrabarti2009transient}
\bibinfo{author}{\bibfnamefont{R.}~\bibnamefont{Chakrabarti}},
  \bibinfo{journal}{Pramana} \textbf{\bibinfo{volume}{72}},
  \bibinfo{pages}{665} (\bibinfo{year}{2009}).

\bibitem[{\citenamefont{Mai and Dhar}(2007)}]{mai2007nonequilibrium}
\bibinfo{author}{\bibfnamefont{T.}~\bibnamefont{Mai}} \bibnamefont{and}
  \bibinfo{author}{\bibfnamefont{A.}~\bibnamefont{Dhar}},
  \bibinfo{journal}{Phys. Rev. E} \textbf{\bibinfo{volume}{75}},
  \bibinfo{pages}{061101} (\bibinfo{year}{2007}).

\bibitem[{\citenamefont{Ciliberto et~al.}(2010)\citenamefont{Ciliberto,
  Joubaud, and Petrosyan}}]{ciliberto2010fluctuations}
\bibinfo{author}{\bibfnamefont{S.}~\bibnamefont{Ciliberto}},
  \bibinfo{author}{\bibfnamefont{S.}~\bibnamefont{Joubaud}}, \bibnamefont{and}
  \bibinfo{author}{\bibfnamefont{A.}~\bibnamefont{Petrosyan}},
  \bibinfo{journal}{J. Stat. Mech} \textbf{\bibinfo{volume}{P12003}}
  (\bibinfo{year}{2010}).

\bibitem[{\citenamefont{Sen et~al.}(2011)\citenamefont{Sen, Baura, and
  Bag}}]{sen2011work}
\bibinfo{author}{\bibfnamefont{M.~K.} \bibnamefont{Sen}},
  \bibinfo{author}{\bibfnamefont{A.}~\bibnamefont{Baura}}, \bibnamefont{and}
  \bibinfo{author}{\bibfnamefont{B.~C.} \bibnamefont{Bag}},
  \bibinfo{journal}{Euro. Phys. J. B} \textbf{\bibinfo{volume}{83}},
  \bibinfo{pages}{381} (\bibinfo{year}{2011}).

\end{thebibliography}

\end{document}